\documentclass[%
 reprint,
 amsmath,amssymb,
 aps,
 prl,
]{revtex4-2}
\usepackage{CJK}
\usepackage{graphicx}
\usepackage{dcolumn}
\usepackage{bm}
\usepackage{xcolor}
\usepackage{amsmath}
\usepackage{float} 
\usepackage{physics}
\usepackage{chemfig}
\usepackage[colorinlistoftodos]{todonotes}
\begin{document}

\preprint{APS/123-QED}

\title{Modeling of Proton Interaction with Organic Polymers:\\Implications for Cancer Therapy and Beyond}

\author{F.~Matias\textsuperscript{1}}
\author{T. F.~Silva\textsuperscript{2}}
\author{N. E.~Koval\textsuperscript{3}}
\author{J. J. N.~Pereira\textsuperscript{1}}
\author{P. C. G.~Antunes\textsuperscript{1}}
\author{P. T. D.~Siqueira\textsuperscript{1}}
\author{M. H.~Tabacniks\textsuperscript{2}}
\author{H.~Yoriyaz\textsuperscript{1}}
\author{J. M. B.~Shorto\textsuperscript{1}}
\author{P.L.~Grande\textsuperscript{4}}

\affiliation{\textsuperscript{1}Instituto de Pesquisas Energ\'eticas e Nucleares, Av. Professor Lineu Prestes, S\~ao Paulo, 05508-000, Brazil}

\affiliation{\textsuperscript{2}\mbox{Instituto de F\'isica da Universidade de S\~ao Paulo, Rua do Mat\~ao, trav. R187, S\~ao Paulo, 05508-090, Brazil}}

\affiliation{\textsuperscript{3}Centro de F\'isica de Materiales, Paseo Manuel de Lardizabal 5, 20018 Donostia-San Sebasti\'an, Spain}


\affiliation{\textsuperscript{4}\mbox{Instituto de F\'isica da Universidade Federal do Rio Grande do Sul, Av. Bento Gon\c{c}alves, Porto Alegre, 9500, Brazil}}



\date{\today}

\begin{abstract}

\noindent
This comprehensive study delves into the intricate interplay between protons and organic polymers, offering insights into proton therapy in cancer treatment. Focusing on the influence of the spatial electron density distribution on stopping power estimates, we employed time-dependent density functional theory (TDDFT), coupled with the Penn method. Surprisingly, the assumption of electron density homogeneity in polymers is fundamentally flawed, resulting in an overestimation of stopping power values at energies below 2 MeV, approximately. Moreover, Bragg's rule application in specific compounds exhibited significant deviations from experimental data in the Bragg peak region, challenging established norms.
\end{abstract}

\maketitle


In the last two decades, clinical therapy using proton beams for the treatment of cancerous tumors has experienced steady growth \cite{delaney2005,bernier2004,baskar2012,leticia2022}. Although this form of radiotherapy already counts with highly developed technology, it still retains significant challenges in terms of physical and clinical aspects \cite{loeffler2013,durante2017,durante2021}. One of these challenges is the precise accounting of relative biological effectiveness (RBE), which is the ratio between the doses required by two types of radiation to cause the same biological effect. This factor, measurable through Linear energy transfer (LET) or microdosimetry, depends on how the energy is deposited on a micrometric scale \cite{grassberger2011}.

In proton therapy, RBE is traditionally defined by a constant value of 1.1 (relative to X-ray dose) for all points along the beam path and all stopping points \cite{paganetti2002,paganetti2019}. However, a comprehensive review of the available experimental data in the literature \cite{paganetti2014} reveals that, despite a lack of experimental standardization and, large uncertainties, there is evidence that RBE values vary considerably and can exceed 1.1 at the end of the beam range. These differences have clinical implications \cite{peeler2016,underwood2017}, therefore, it is of great importance to reduce experimental uncertainties to accurately describe the effects of proton beams on tissues.

The energy of the proton transferred to the biological tissue is directly related to its velocity. As the proton slows down, the amount of energy transferred to the tissue increases, resulting in maximum dose deposition at a specific depth. This region around the peak of maximum dose deposition is known as the Bragg peak \cite{bragg1905,william1946,bragg:gordon,brown2004}. It is the region of greatest interest in proton beam radiotherapy applications, and its precise positioning is crucial during the definition of the irradiation plan. This particular profile of proton beam energy deposition presents significant clinical advantages, especially for pediatric patients, by allowing optimal dose delivery to tumor tissue and by minimizing dose to organs at risk in surrounding areas, thus reducing the chances of future complications and induction of secondary tumors \cite{hall2006,merchant2009,zhang2014}.

On the other hand, the high and relatively narrow dose peak makes both quality control in dose monitoring, and precise patient positioning, even more crucial, with the risk of damaging health tissues with high radiation dose. Therefore, in-depth investigations of the uncertainties in the range and stopping power values are essential for a more accurate dose distribution in patients \cite{ma2012,paganetti2012}.

The accurate knowledge of electronic stopping power, or electronic stopping cross-section (SCS), is not just essential in proton therapy. It is also important in many fields of science and technological applications, such as outer space exploration (space weathering), nanotechnology (ion beam modeling), material modifications, and nuclear fusion research (plasma-wall interaction) \cite{Brunetto,Kim,Papale:2015,Meisl,mayer_ion_2019}. However, as depicted above, its most critical application lies in dosimetry for cancer treatment using ions, given the increasing global use of protons and heavier ions in radiation therapy and the risks involved \cite{Newhauser:2015,Durante}. Therefore, SCS is a fundamental quantity that requires a detailed understanding of energy-loss processes.

In the realm of theoretical investigations into ion-matter interactions, it is customary to employ simplified models, exemplified by the homogeneous free electron gas (FEG) model, for the representation of valence electrons within materials. This pragmatic approach facilitates straightforward predictions of stopping power and yields results that closely align with experimental data \cite{pruneda2007,quijada2007,goebl2014,roth2017a,roth2017b,sortica2017,matias2017,Matias:2019,ullah2018}.

Although the FEG model is reliable for materials with simple electronic structures, its effectiveness diminishes when dealing with materials characterized by complex electronic excitations. Here, we demonstrate that these materials can still be treated as a collection of FEG with high accuracy, ensuring simplicity and avoiding time-consuming full atomistic \emph{ab initio} calculations.

For this purpose, we utilized stopping power for a FEG with different densities or plasmon frequencies from the calculations of the time-dependent density functional theory (TDDFT) \cite{quijada2007,koval2013,Koval2017,matias2017}. The results were averaged according to the Penn method \cite{Maarten:2019b}. 

In this framework, knowledge of the energy-loss function (ELF) of materials is essential. The Penn approach \cite{penn1987} introduced an algorithm to determine the electron inelastic mean free paths (IMFP) by utilizing a model dielectric function derived from the experimental ELF specific to the material under investigation. The same model has been applied to estimate the electron stopping power in various materials \cite{SHINOTSUKA201275} and has been extended to calculate the non-linear stopping power of ions \cite{Maarten:2019b}. This extension involves using the ELF to appropriately weight contributions from different electron gas components within a statistical ensemble that characterizes the material of interest. ELF functions at the optical limit can be found for different materials elsewhere \cite{ding:ELF}.

For compounds such as hydrocarbons, Bragg's rule has been used to calculate the stopping values of their constituent elements \cite{bragg1905}. However, it can cause significant differences in stopping power in the Bragg peak region compared to the experimental data \cite{lodhi1974}.

By applying the proposed formalism, we examined the validity of Bragg's rule in the context of organic polymers. Specifically, we examined the cases of polyethylene (PE), polystyrene (PS), poly(2-vinylpyridine) (P2VP), polyacetylene (PA), poly(methyl methacrylate) (PMMA) and polyimide (PI). The study of these polymers is important because virtually all phantoms used for dose verification and quality assurance in proton therapy treatments are manufactured with polymers such as PMMA. Furthermore, some components that make up the proton accelerators are constructed with PE or PS \cite{kirby2010,deVera2013,christina2015,battaglia2016,casolaro2019,rezaeian2023}.

TDDFT is a highly effective \emph{ab initio} tool for describing electronic stopping power in spherical jelliums. The jellium model assumes a positive background (representing the ion cores) that provides a charge balancing for the electron gas. The advantage of such representation as compared to fully atomistic models is the computational efficiency. The TDDFT in a FEG has been shown to provide accurate results for near-free-electron systems whereas using an atomistic representation requires careful consideration of trajectories to calculate the random stopping power \cite{gu2020, kononov2023}.

The positive background density of the jellium with radius $R_{\rm cl}$ is defined by $n_{0}^{+}({\bf r})=n_{0}^{+} \Theta (R_{\rm-r cl})$, where $\Theta(x)$ denotes the Heaviside function. The electronic density of the cluster is determined by the Wigner-Seitz radius, $r_{s}$ ($4\pi r_{s}^{3}/3=1/n_{0}$). The total number of electrons in the cluster, $N_{e}$, is then given by $N_{e}=(R_{cl}/r_{s})^{3}$.
 
Although there have been minor refinements in terms of accuracy, the approach
adopted in this work reflects the methodology used in \cite{quijada2007,koval12,koval2013,matias2017}, and as such it will be briefly explained in this section. In this approach, the time evolution of electronic density incorporates, in a non-perturbative manner, the complete dynamic interaction between an external field and the medium. This computational framework has been used to analyze various issues in condensed matter systems, such as dynamic charge screening in metallic media \cite{borisov2004}, energy loss of atomic particles in matter \cite{pruneda2007,quijada2007,matias2017}, as well as many-body effects associated with hole screening in photoemission \cite{koval12}.

A static density functional theory (DFT) calculation is performed to obtain the system's ground state. The time evolution of the complete electronic density, $n({\bf r},t)$, in response to an external field (in this case, a proton), is conducted within the framework of TDDFT in the Kohn-Sham regime (KS-TDDFT) (atomic units are used throughout, unless specified otherwise):

\begin{equation}
i\frac{\partial \psi _{j}({\bf r},t)}{\partial t} ~=~\left\{
T+V_{\rm{eff}}([n],{\bf r},t)\right\} \psi _{j}({\bf
r},t)~\mathrm{,} \label{kseq}
\end{equation}

\noindent
where $\psi_j({\bf r},t)$ are the Kohn-Sham orbitals and $T$ is the kinetic energy operator. The Kohn-Sham effective potential, $V_{\text{eff}}([n],{\bf r},t)$, is a function of the electronic density of the system:
$n({\bf r},t)= \sum_{j\in occ.}{\left| \psi_{j}({\bf r},t)\right|
^{2}}$. $V_{\text{eff}}$ is obtained as the sum of the external potential $V_{\text{ext}}^+$, the Hartree potential $V_{\text{H}}$, and the exchange-correlation potential $V_{\text{xc}}$:
$V_{\rm{eff}}=V_{\rm{ext}}^++V_{\rm{H}}+V_{\rm{xc}}$.
$V_{\text{ext}}^+$ is the potential created by the proton. $V_{\text{xc}}({\bf r},t)$ is treated within a standard adiabatic local density approximation (ALDA) approach.
The numerical procedure is very similar to that employed in Refs. \cite{borisov2004,pruneda2007,quijada2007,quijada10}, where additional details can be found.

The energy loss is calculated by integrating the time-dependent induced force over the proton:

\begin{equation}
E_{\rm{loss}}=-v\int F_z (t)dt,
\end{equation}

\noindent
where $v$ is the (constant) velocity at which the proton traverses the jellium. Once the induced force on the proton is calculated, the average or effective stopping power is computed as the energy loss per unit path length, i.e.

\begin{equation}
\left(\frac{dE}{dz}\right)_{\text{TDDFT}}=\frac{E_{\rm{loss}}}{2R_{cl}}.
\label{stp_tddft}
\end{equation}

Recently, an alternative non-linear method has been introduced to characterize the stopping power of light and heavy ions in materials \cite{Maarten:2019b}. This method incorporates the influence of non-free electron distributions within a theoretical model for stopping power calculations, such as TDDFT. For a low energy proton ($v < v_F$, where $v_F$ is Fermi velocity), the Penn approach has been used recently in the transport cross section (TCS) \cite{Matias:2019}. This approach considers the combination of electron-gas responses characterized by nonuniform densities, similar to the approach outlined in the Penn method \cite{penn1987}. 

To achieve this goal, each free electron density is analyzed based on the material's ELF at the optical limit, as follows:
\begin{eqnarray}
       g(\omega_p) =\frac{2}{\pi \omega_p}\text{ELF}(\omega_p).
    \label{eq:elf} 
\end{eqnarray}
\noindent
The stopping power depends on the plasmon frequency $\omega_p$, a value determined by the individual electron gas contributions obtained from $r_s$. $\omega_p=\sqrt{3}r_s^{-3/2}$. Therefore, the stopping power is now calculated as follows:

\begin{equation}
    \left(\frac{dE}{dz}\right)_{\text{TDDFT-Penn}} = \int_0^{\infty}d\omega_p g(\omega_p)\left(\frac{dE}{dz}\right)_{\text{TDDFT}}(\omega_p,v).
    \label{eq:penn}
\end{equation}

\noindent
In the above equation, the term $\left(dE/dz\right)_{\text{TDDFT}}$ is calculated in the TDDFT framework using Equation (\ref{stp_tddft}). Because of that, we named this approach TDDFT-Penn.

\begin{figure}[H]
\begin{center}
\includegraphics[scale=0.36,clip]{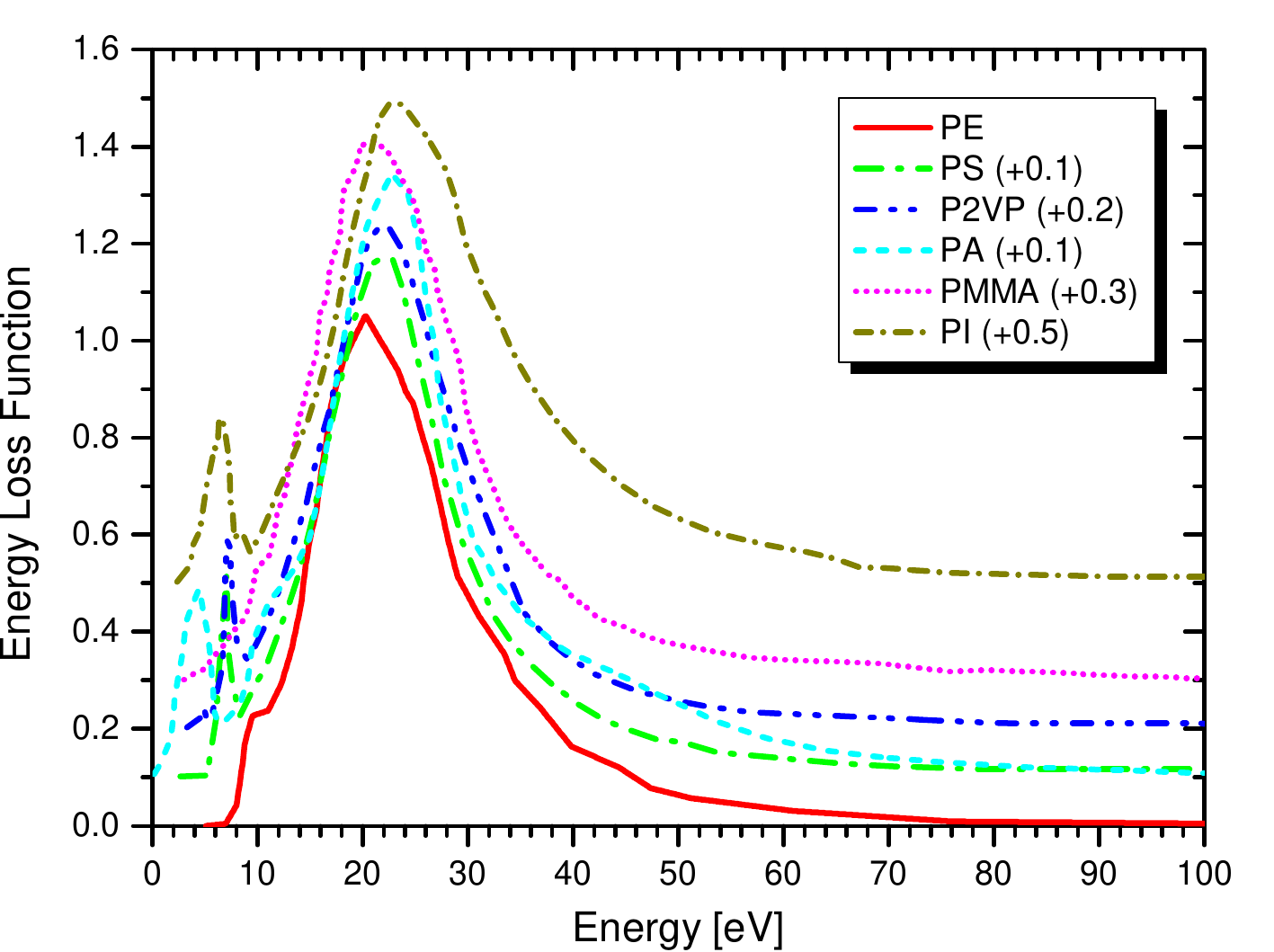}
\caption{Optical-ELF data for PE, PS, P2VP, PA, PMMA, and PI obtained from \cite{tanuma1994a,inagaki1977,deVera2011} and used to calculate electronic SCS with the TDDFT-Penn approach.}
\label{fig1}
\end{center}
\end{figure}

\begin{figure}[H]
\begin{center}
\includegraphics[scale=0.36,clip]{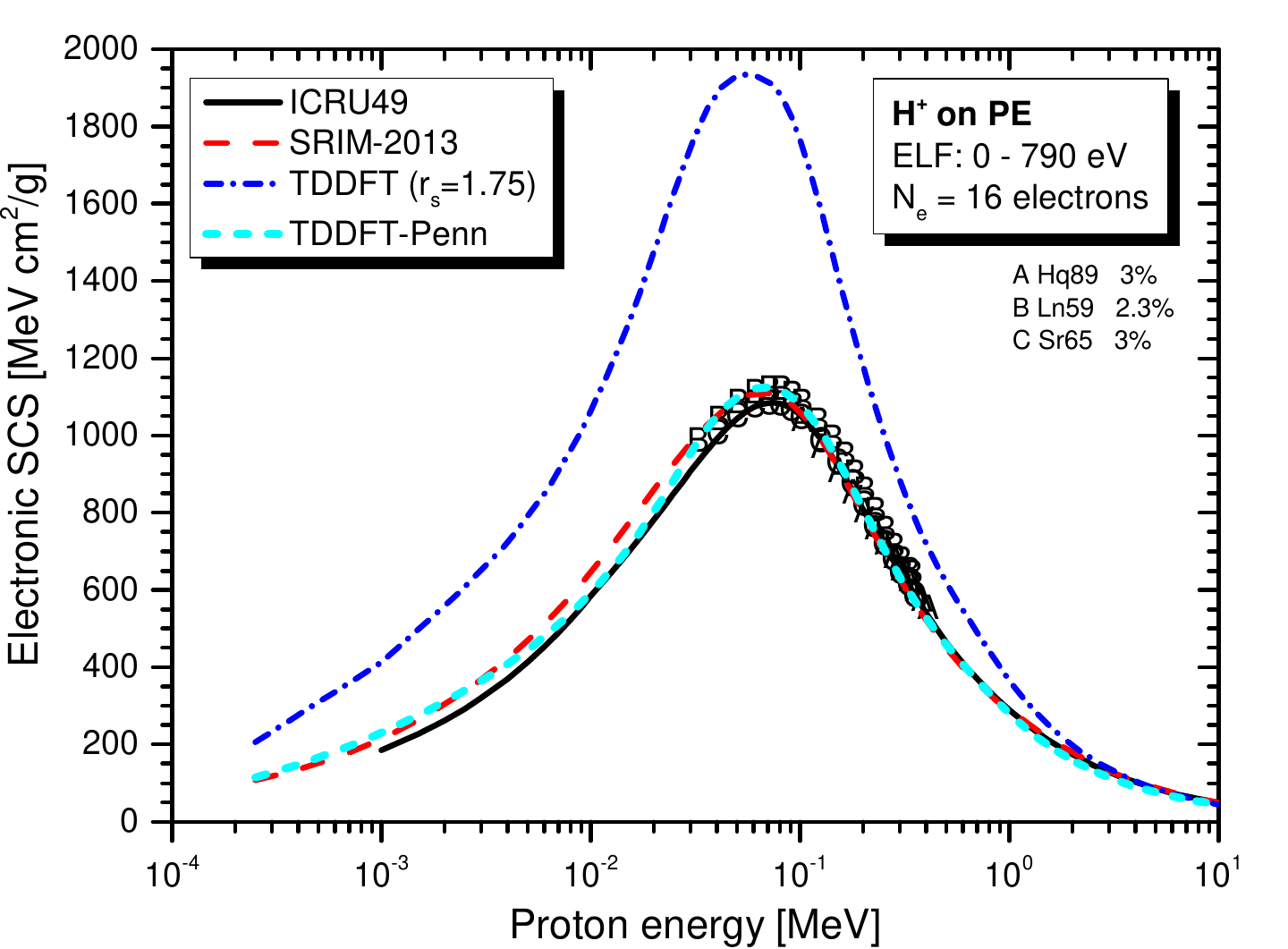}
\caption{Proton SCS in PE polymer. TDDFT results with a unique FEG ($r_s=1.75$ au) are shown in the blue dash-dot line, and the TDDFT-Penn is in the cyan short dash line. Experimental data (uppercase letters) at Bragg peak \cite{iaea}. Semi-empirical models ICRU49 \cite{icru49:1993} and SRIM-2013 \cite{srim2010} presented.}
\label{fig2}
\end{center}
\end{figure}

\begin{figure}[H]
\begin{center}
\includegraphics[scale=0.36,]{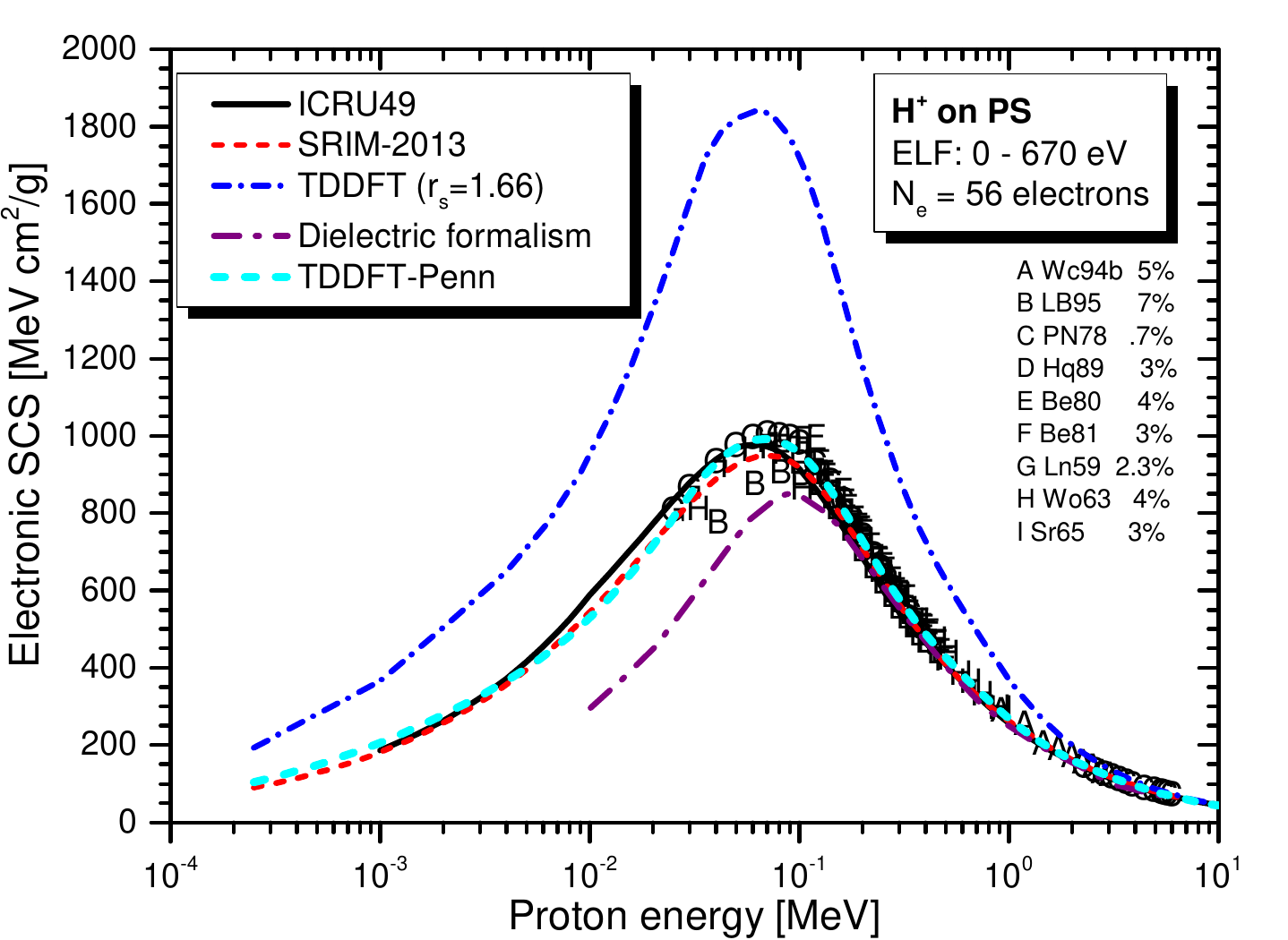}
\caption{Proton SCS in PS polymer. TDDFT results using a unique FEG ($r_s=1.66$ au) and TDDFT-Penn. Experimental data (uppercase letters) concentrated at Bragg peak. Dielectric formalism results in purple dash-dot line \cite{deVera2013}. Semi-empirical models ICRU49 \cite{icru49:1993} and SRIM-2013 \cite{srim2010} showcased.}
\label{fig3}
\end{center}
\end{figure}

\begin{figure}[H]
\begin{center}
\includegraphics[scale=0.36,clip]{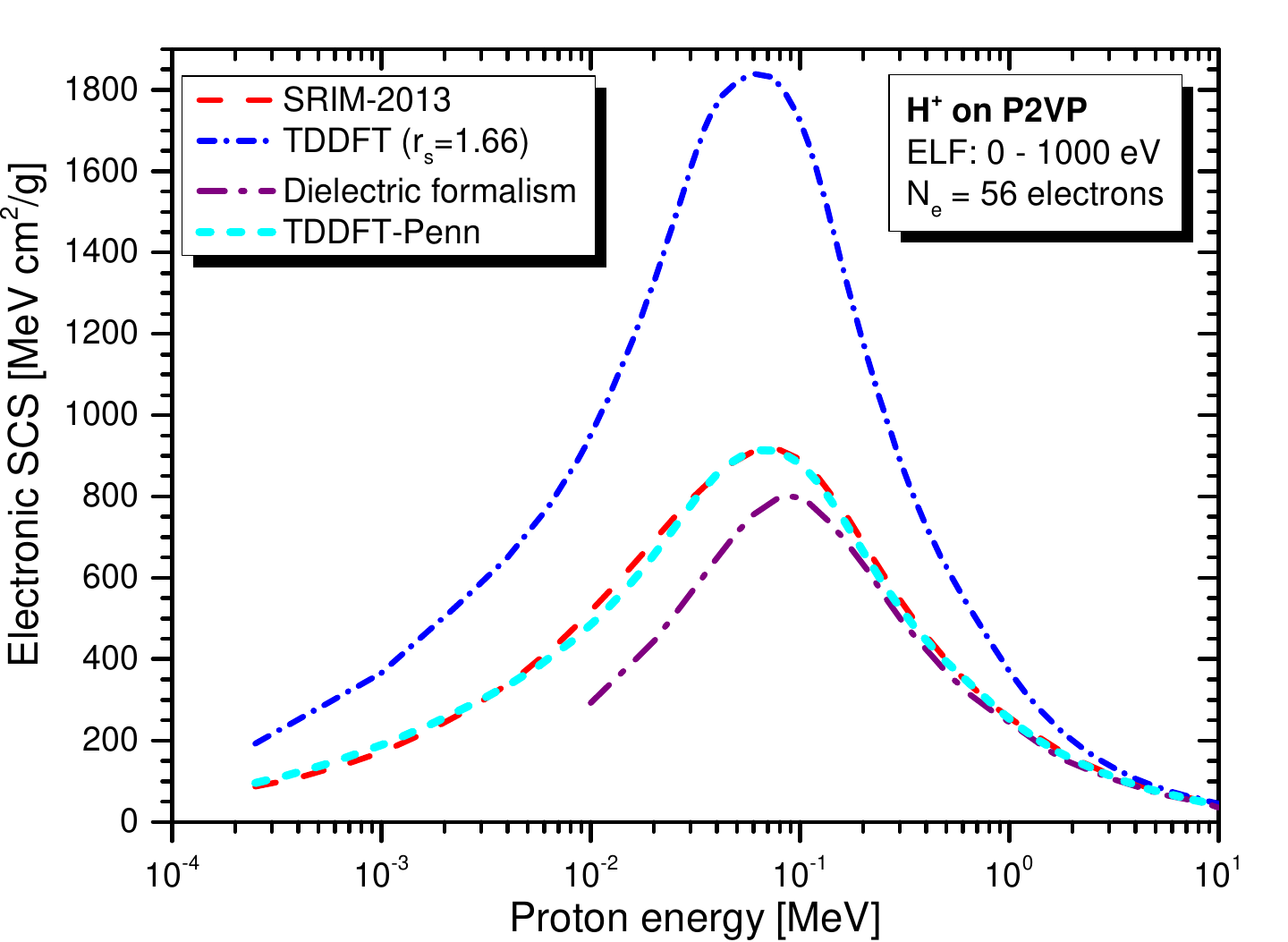}
\caption{Proton SCS in P2VP polymer. TDDFT results with a unique FEG ($r_s=1.66$ au) and TDDFT-Penn. Results based on dielectric formalism \cite{deVera2011}. Semi-empirical model SRIM-2013 \cite{srim2010} included.}
\label{fig4}
\end{center}
\end{figure}

\begin{figure}[H]
\begin{center}
\includegraphics[scale=0.36,clip]{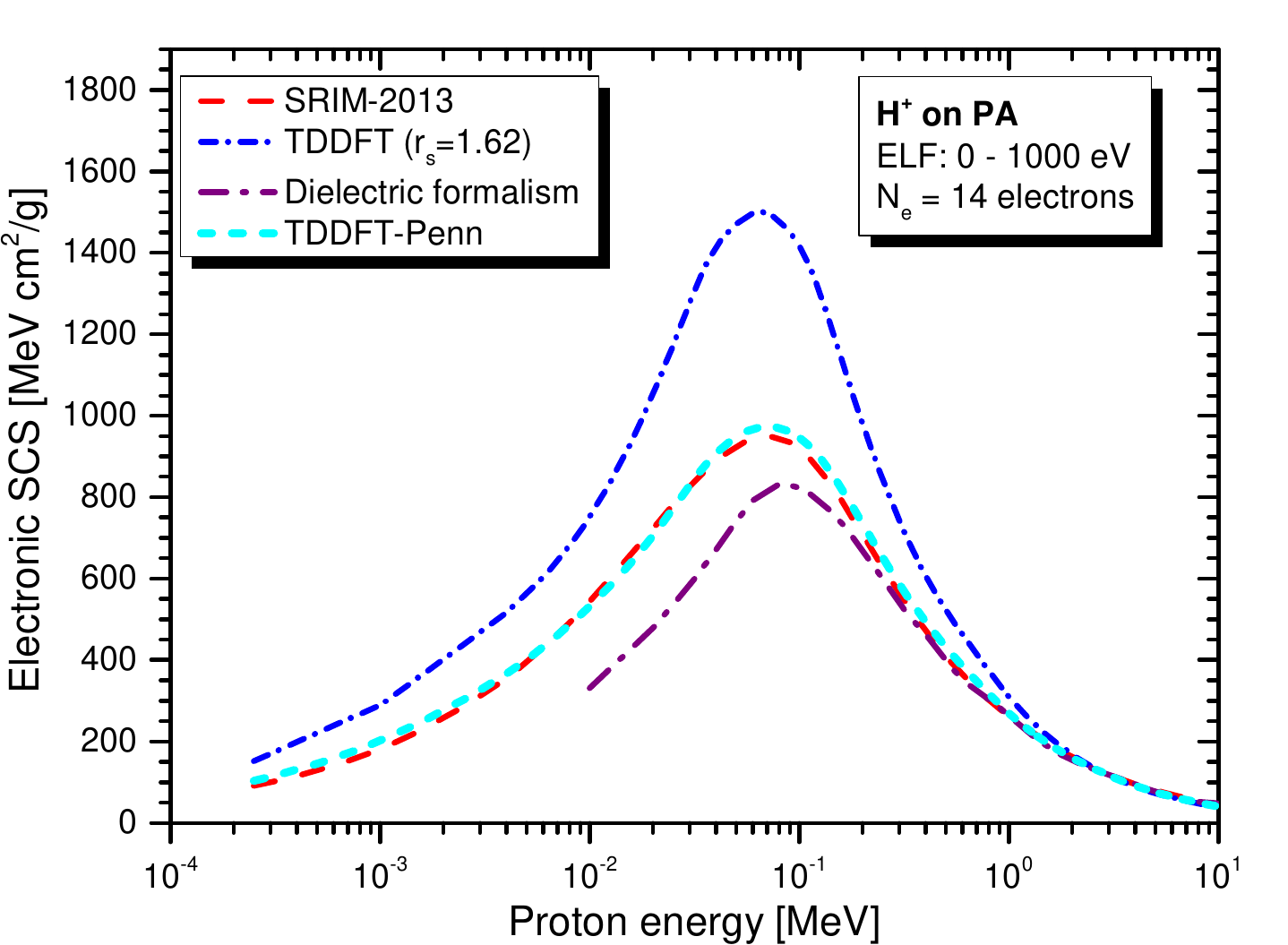}
\caption{Proton SCS in PA polymer. TDDFT results with a unique FEG ($r_s=1.62$ au) and TDDFT-Penn. Results based on dielectric formalism \cite{deVera2011}. Semi-empirical model SRIM-2013 \cite{srim2010} presented.}
\label{fig5}
\end{center}
\end{figure}

\begin{figure}[H]
\begin{center}
\includegraphics[scale=0.36,clip]{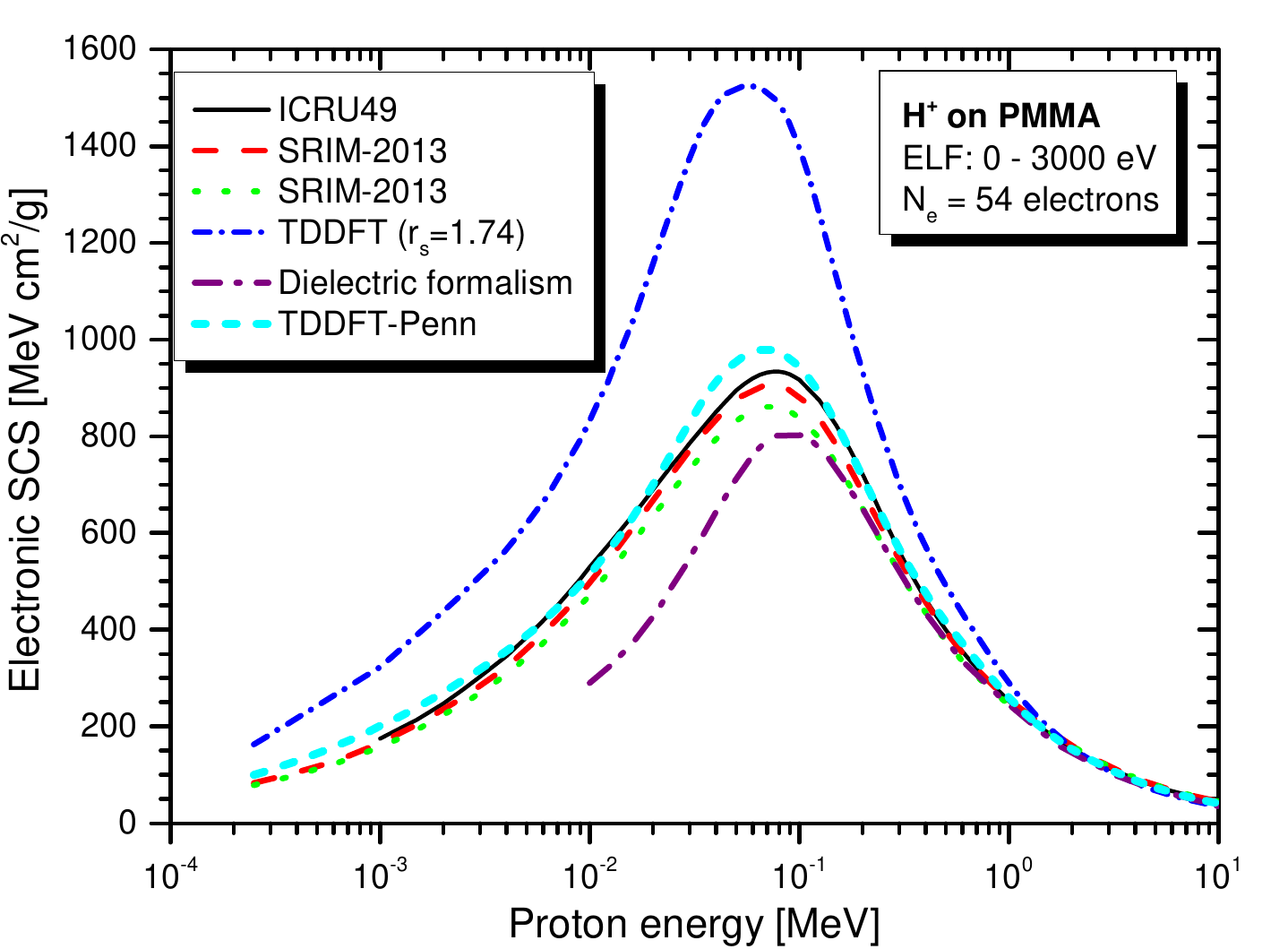}
\caption{Proton SCS in PMMA polymer. TDDFT results with a unique FEG ($r_s=1.74$ au) and TDDFT-Penn. Dielectric formalism results \cite{deVera2011}. Semi-empirical models ICRU49 \cite{icru49:1993} and SRIM-2013 \cite{srim2010} with Bragg's rule (red dashed line) and CAB (dotted green line) correction showcased.}
\label{fig6}
\end{center}
\end{figure}

\begin{figure}[H]
\begin{center}
\includegraphics[scale=0.36,clip]{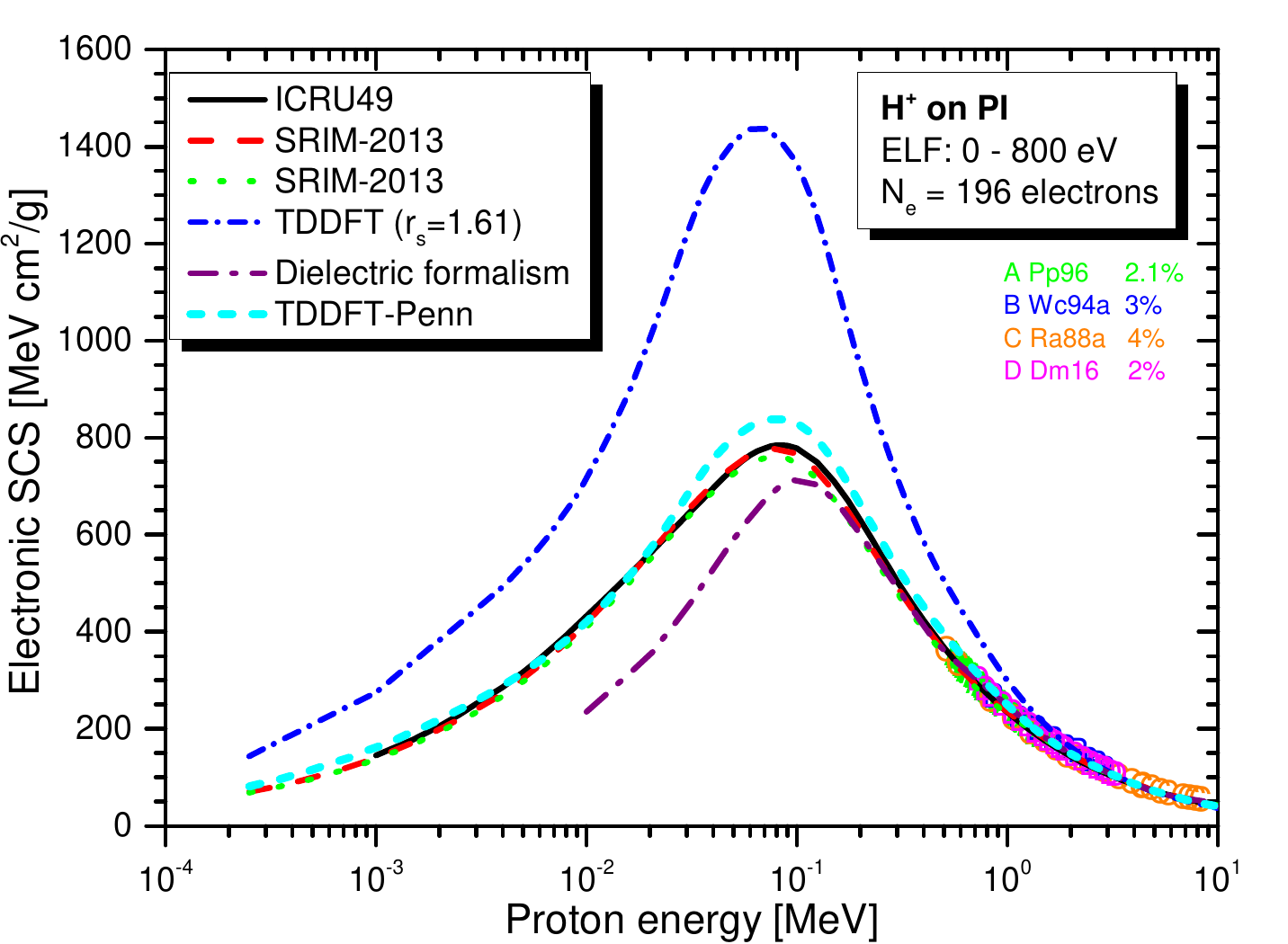}
\caption{Proton SCS in PI polymer. TDDFT results with a unique FEG ($r_s=1.61$ au) and TDDFT-Penn. Results based on dielectric formalism \cite{deVera2011}. Semi-empirical models ICRU37 \cite{icru37:1984} and SRIM-2013 \cite{srim2010} with Bragg's rule and CAB correction presented.}
\label{fig7}
\end{center}
\end{figure}

TDDFT-Penn calculations were performed according to Equations (\ref{stp_tddft}) to (\ref{eq:penn}), and the electronic SCS results for PE, PS, P2VP, PA, PMMA, and PI to energetic protons are presented in Figure~\ref{fig2} to \ref{fig7}, respectively. The data used to calculate SCS with the TDDFT-Penn method are listed in Table~\ref{tab:polymer_data} and the optical-ELF data for each polymer are shown in Figure~\ref{fig1}. 

\begin{table}[H]
\centering
\caption{Data used in the TDDFT-Penn approach to calculate the electronic SCS of different polymers based on their monomers \cite{tanuma1994a,inagaki1977,deVera2011}.}
\begin{tabular}{ccccc}
\hline\hline
Polymer & \begin{tabular}[c]{@{}c@{}}Formula\end{tabular} & \begin{tabular}[c]{@{}c@{}}ELF\\ range (eV)\end{tabular} & \begin{tabular}[c]{@{}c@{}}Total/valence \\ electrons\end{tabular} & \begin{tabular}[c]{@{}c@{}} $\rho$\\ (g/cm$^3$) \end{tabular} \\ \hline
PE      & (C$_2$H$_4$)$_{\rm n}$ &  0-790 &  16/12 & 0.93\\
PS      & (C$_8$H$_8$)$_{\rm n}$ &  0-670 &  56/40 & 1.06\\
P2VP    & (C$_7$H$_7$N)$_{\rm n}$ & 0-1000 &  56/40 & 1.15\\
PA      & (C$_2$H$_2$)$_{\rm n}$ & 0-1000 &  14/10 & 1.36\\
PMMA    & (C$_8$H$_8$O$_2$)$_{\rm n}$ & 0-3000 &  54/40 & 1.19\\
PI      & (C$_{22}$H$_{10}$N$_2$O$_5$)$_{\rm n}$ & 0-800  & 196/138 & 1.42\\ \hline\hline
\end{tabular}
\label{tab:polymer_data}
\end{table}

We compare the results of our approach with ICRU49 \cite{icru49:1993}, ICRU37 \cite{icru37:1984}, SRIM-2013 \cite{srim2010}, and TDDFT with the homogeneous assumption. This comparison shows the need to completely break down the assumption of spatial homogeneity of the valence electron density in complex materials, such as polymers. For example, the homogeneous assumption leads to an overestimation of the SCS values for proton energies below approximately 2000 keV, compared to the SRIM-2013 data. 

\noindent
Our approach produces more realistic values and, at the same time, offers a physically sound approach to dealing with inhomogeneities. 

We also included in the comparison the experimental data available at the IAEA database \cite{iaea} (upper case letters), with which the TDDFT-Penn results show an excellent agreement, as well as with semi-empirical SRIM-2013 \cite{srim2010} results (red dashed line) using Bragg's rule \cite{bragg1905}, as can be seen in Figures~\ref{fig2} and \ref{fig3}. 

Finally, SCS results from the dielectric formalism, particularly the Mermin-Energy-Loss-Function Generalized Oscillator Strength model (MELF-GOS) \cite{deVera2011}, are also included in the comparisons. This approach also considers inhomogeneities in the electron density of the material utilizing a similar experimental ELF and thus both methods will give similar mean excitation energies (I) (occurring in the Bethe formula for fast projectiles). However, it underestimates the SCS at low energies by a significant amount. Several studies use this approach to calculate SCS in biological media \cite{deVera2011,deVera2013,molina2013}. Because this theoretical model is linear, it loses accuracy for ion energies in the Bragg peak region. In this energy range, the non-linear effects become significant. Even though we have not presented MELF-GOS results for PE, we expect similar behavior as observed in the others.

In particular, the TDDFT-Penn results for PS (refer to Figure~\ref{fig3}) agree better with the experimental data compared to SRIM-2013. SRIM-2013 employs Bragg's rule, resulting in an excitation energy (I) for PS of 65.5 eV \cite{dalton1968}. However, electron energy loss spectroscopy (EELS) data from experiments \cite{inagaki1977} for the compound PS suggest a lower I value (59.3 eV). This observation explains the higher values obtained with our approach in the Bragg peak region.

SCS results for P2VP and PA (see Figures~\ref{fig4} and \ref{fig5}) agree well with SRIM-2013. However, it is important to note that we did not find any experimental data for comparison in this context. Furthermore, the dielectric approach, employing a similar ELF function as input \cite{deVera2011}, deviates significantly below the Bragg peak.

Interestingly, for PMMA and PI (see Figures~\ref{fig6} and \ref{fig7}) Bragg's rule indicates 74 eV \cite{icru49:1993} and 79.6 eV \cite{icru37:1984}, while the experimental ELF \cite{deVera2011} points to significantly lower values of 66 eV and 68 eV, respectively. It is worth noting that the core and bond (CAB) correction on SRIM-2013 \cite{srim2010} is small but makes the deviation from our approach even higher pointing to a correction in the opposite direction.

As can be observed in Figures~\ref{fig2} to \ref{fig7} the differences between TDDFT-Penn and SRIM-2003 are small, but more pronounced for PMMA and PI at the Bragg peak. They can be attributed to a stronger breakdown of Bragg's rule as a result of the molecular structures of PMMA and PI, both of which feature bonds between C and O, as well as N.

In contrast to other depicted polymers, these two have particularities in their chemical structures.

\noindent
While PMMA has double \chemfig{C=O} and single \chemfig{C-O} bonds, PI has four sets of \chemfig{C=O} and \chemfig{C-O} bonds, as well as two pairs of \chemfig{N-C} bonds. The electronegativity of the atoms in these bonds varies, with oxygen being more electronegative than nitrogen, which is more electronegative than carbon.

Analogously, in TiN compounds, there is a transfer of 1.51 electrons from titanium to nitrogen \cite{Matias:2019}, and the transferred charges. The transfer of charges is expected to be more noticeable in the double bonds between carbon and oxygen in polymers like PMMA and PI. Therefore, our results suggest the possibility of charge transfer occurring from carbon to oxygen or nitrogen on these polymers. If this transfer is indeed occurring, employing Bragg's rule will likely lead to a decreased accuracy in predicting the SCS. Indeed, the SRIM-2013 results (red dashed line) for the PMMA and PI compounds are numerically lower in the Bragg peak region compared to TDDFT-Penn predictions.

Decades of theoretical efforts to model electronic SCS processes have led to various methods. Despite successes, current approaches fall short of uniformly covering a wide energy range. Some models focus on specific energy ranges, while others achieve agreement by treating inner and valence electrons differently. Our \emph{ab initio} framework, using various FEG models and the Penn method, shows excellent agreement across a broad energy spectrum for all electrons.

Exploring proton-stopping power in organic polymers, we highlight the challenge at the Bragg peak for traditional single-FEG models. Our method effectively addresses electron density variations, which shows agreement with experimental data and reference tables for various polymers. The importance of considering intricate electronic structures in theoretical stopping power modeling for polymers is underscored.

The accurate prediction of stopping power requires the inclusion of complex electronic structures. The TDDFT-Penn approach's agreement with experimental data underscores the precision of our theoretical framework, necessitating further validation, particularly for polymers with limited data.

Discrepancies between the TDDFT-Penn approach and semi-empirical models, such as SRIM-2013, underscore the potential impact of molecular structure on predictions. Varied electronegativities in PMMA and PI bonds contribute to charge transfer effects not fully accounted for by Bragg's rule. The primary consequence is a 10\% reduction in mean excitation energy and a 3 mm decrease in the projectile range for 200 MeV protons.

In conclusion, this research challenges assumptions, emphasizing precise modeling for materials with complex electronic structures. Enhancing understanding of ion-polymer interactions, this study provides a robust foundation for applications in proton therapy and fields requiring accurate stopping power predictions.

\begin{acknowledgments}
 This work was partially supported by IPEN (project number 2020.06.IPEN.32) and CNPq (project number 406982/2021-0). The authors acknowledge FAPESP for supporting the computer cluster (process numbers 2012/04583-8 and 2020/04867-2). TFS and MHT acknowledge the financial support provided by CNPq-INCT-FNA (project number 464898/2014-5).
\end{acknowledgments}




\bibliography{apssamp}

\providecommand{\noopsort}[1]{}\providecommand{\singleletter}[1]{#1}%
\begin{thebibliography}{65}%
\makeatletter
\providecommand \@ifxundefined [1]{%
 \@ifx{#1\undefined}
}%
\providecommand \@ifnum [1]{%
 \ifnum #1\expandafter \@firstoftwo
 \else \expandafter \@secondoftwo
 \fi
}%
\providecommand \@ifx [1]{%
 \ifx #1\expandafter \@firstoftwo
 \else \expandafter \@secondoftwo
 \fi
}%
\providecommand \natexlab [1]{#1}%
\providecommand \enquote  [1]{``#1''}%
\providecommand \bibnamefont  [1]{#1}%
\providecommand \bibfnamefont [1]{#1}%
\providecommand \citenamefont [1]{#1}%
\providecommand \href@noop [0]{\@secondoftwo}%
\providecommand \href [0]{\begingroup \@sanitize@url \@href}%
\providecommand \@href[1]{\@@startlink{#1}\@@href}%
\providecommand \@@href[1]{\endgroup#1\@@endlink}%
\providecommand \@sanitize@url [0]{\catcode `\\12\catcode `\$12\catcode
  `\&12\catcode `\#12\catcode `\^12\catcode `\_12\catcode `\%12\relax}%
\providecommand \@@startlink[1]{}%
\providecommand \@@endlink[0]{}%
\providecommand \url  [0]{\begingroup\@sanitize@url \@url }%
\providecommand \@url [1]{\endgroup\@href {#1}{\urlprefix }}%
\providecommand \urlprefix  [0]{URL }%
\providecommand \Eprint [0]{\href }%
\providecommand \doibase [0]{https://doi.org/}%
\providecommand \selectlanguage [0]{\@gobble}%
\providecommand \bibinfo  [0]{\@secondoftwo}%
\providecommand \bibfield  [0]{\@secondoftwo}%
\providecommand \translation [1]{[#1]}%
\providecommand \BibitemOpen [0]{}%
\providecommand \bibitemStop [0]{}%
\providecommand \bibitemNoStop [0]{.\EOS\space}%
\providecommand \EOS [0]{\spacefactor3000\relax}%
\providecommand \BibitemShut  [1]{\csname bibitem#1\endcsname}%
\let\auto@bib@innerbib\@empty
\bibitem [{\citenamefont {Delaney}\ \emph {et~al.}(p 15)\citenamefont
  {Delaney}, \citenamefont {Jacob}, \citenamefont {Featherstone},\ and\
  \citenamefont {Barton}}]{delaney2005}%
  \BibitemOpen
  \bibfield  {author} {\bibinfo {author} {\bibfnamefont {G.}~\bibnamefont
  {Delaney}}, \bibinfo {author} {\bibfnamefont {S.}~\bibnamefont {Jacob}},
  \bibinfo {author} {\bibfnamefont {C.}~\bibnamefont {Featherstone}},\ and\
  \bibinfo {author} {\bibfnamefont {M.}~\bibnamefont {Barton}},\ }\bibfield
  {title} {\bibinfo {title} {The role of radiotherapy in cancer treatment:
  estimating optimal utilization from a review of evidence-based clinical
  guidelines},\ }\href@noop {} {\bibfield  {journal} {\bibinfo  {journal}
  {Cancer}\ }\textbf {\bibinfo {volume} {104(6)}},\ \bibinfo {pages} {1129}
  (\bibinfo {year} {2005 Sep 15})}\BibitemShut {NoStop}%
\bibitem [{\citenamefont {Bernier}\ \emph {et~al.}( Sep)\citenamefont
  {Bernier}, \citenamefont {Hall},\ and\ \citenamefont
  {Giaccia}}]{bernier2004}%
  \BibitemOpen
  \bibfield  {author} {\bibinfo {author} {\bibfnamefont {J.}~\bibnamefont
  {Bernier}}, \bibinfo {author} {\bibfnamefont {E.~J.}\ \bibnamefont {Hall}},\
  and\ \bibinfo {author} {\bibfnamefont {A.}~\bibnamefont {Giaccia}},\
  }\bibfield  {title} {\bibinfo {title} {Radiation oncology: a century of
  achievements},\ }\href@noop {} {\bibfield  {journal} {\bibinfo  {journal}
  {Nature Reviews Cancer}\ }\textbf {\bibinfo {volume} {4(9)}},\ \bibinfo
  {pages} {737–747} (\bibinfo {year} {2004 Sep})}\BibitemShut {NoStop}%
\bibitem [{\citenamefont {Baskar}\ \emph {et~al.}(2012)\citenamefont {Baskar},
  \citenamefont {Lee}, \citenamefont {Yeo},\ and\ \citenamefont
  {Yeoh}}]{baskar2012}%
  \BibitemOpen
  \bibfield  {author} {\bibinfo {author} {\bibfnamefont {R.}~\bibnamefont
  {Baskar}}, \bibinfo {author} {\bibfnamefont {K.}~\bibnamefont {Lee}},
  \bibinfo {author} {\bibfnamefont {R.}~\bibnamefont {Yeo}},\ and\ \bibinfo
  {author} {\bibfnamefont {K.}~\bibnamefont {Yeoh}},\ }\bibfield  {title}
  {\bibinfo {title} {{Cancer and Radiation Therapy: Current Advances and Future
  Directions}},\ }\href@noop {} {\bibfield  {journal} {\bibinfo  {journal} {Int
  J Med Sci.}\ }\textbf {\bibinfo {volume} {9(3)}},\ \bibinfo {pages} {193}
  (\bibinfo {year} {2012})}\BibitemShut {NoStop}%
\bibitem [{\citenamefont {Nogueira}\ \emph {et~al.}(2022)\citenamefont
  {Nogueira}, \citenamefont {Jemal}, \citenamefont {Yabroff},\ and\
  \citenamefont {Efstathiou}}]{leticia2022}%
  \BibitemOpen
  \bibfield  {author} {\bibinfo {author} {\bibfnamefont {L.~M.}\ \bibnamefont
  {Nogueira}}, \bibinfo {author} {\bibfnamefont {A.}~\bibnamefont {Jemal}},
  \bibinfo {author} {\bibfnamefont {K.~R.}\ \bibnamefont {Yabroff}},\ and\
  \bibinfo {author} {\bibfnamefont {J.~A.}\ \bibnamefont {Efstathiou}},\
  }\bibfield  {title} {\bibinfo {title} {{Assessment of Proton Beam Therapy Use
  Among Patients With Newly Diagnosed Cancer in the US, 2004-2018}},\
  }\href@noop {} {\bibfield  {journal} {\bibinfo  {journal} {JAMA Netw Open}\
  }\textbf {\bibinfo {volume} {5(4)}},\ \bibinfo {pages} {e229025} (\bibinfo
  {year} {2022})}\BibitemShut {NoStop}%
\bibitem [{\citenamefont {Loeffler}\ and\ \citenamefont {Durante}(
  Jul)}]{loeffler2013}%
  \BibitemOpen
  \bibfield  {author} {\bibinfo {author} {\bibfnamefont {J.~S.}\ \bibnamefont
  {Loeffler}}\ and\ \bibinfo {author} {\bibfnamefont {M.}~\bibnamefont
  {Durante}},\ }\bibfield  {title} {\bibinfo {title} {Charged particle
  therapy--optimization, challenges and future directions},\ }\href@noop {}
  {\bibfield  {journal} {\bibinfo  {journal} {Nat Rev Clin Oncol}\ }\textbf
  {\bibinfo {volume} {10(7)}},\ \bibinfo {pages} {411} (\bibinfo {year} {2013
  Jul})}\BibitemShut {NoStop}%
\bibitem [{\citenamefont {Durante}\ \emph {et~al.}( Aug)\citenamefont
  {Durante}, \citenamefont {Orecchia},\ and\ \citenamefont
  {Loeffler}}]{durante2017}%
  \BibitemOpen
  \bibfield  {author} {\bibinfo {author} {\bibfnamefont {M.}~\bibnamefont
  {Durante}}, \bibinfo {author} {\bibfnamefont {R.}~\bibnamefont {Orecchia}},\
  and\ \bibinfo {author} {\bibfnamefont {J.~S.}\ \bibnamefont {Loeffler}},\
  }\bibfield  {title} {\bibinfo {title} {Charged-particle therapy in cancer:
  clinical uses and future perspectives},\ }\href@noop {} {\bibfield  {journal}
  {\bibinfo  {journal} {Nat Rev Clin Oncol}\ }\textbf {\bibinfo {volume}
  {14(8)}},\ \bibinfo {pages} {483} (\bibinfo {year} {2017 Aug})}\BibitemShut
  {NoStop}%
\bibitem [{\citenamefont {Durante}\ \emph {et~al.}( Dec)\citenamefont
  {Durante}, \citenamefont {Debus},\ and\ \citenamefont
  {Loeffler}}]{durante2021}%
  \BibitemOpen
  \bibfield  {author} {\bibinfo {author} {\bibfnamefont {M.}~\bibnamefont
  {Durante}}, \bibinfo {author} {\bibfnamefont {J.}~\bibnamefont {Debus}},\
  and\ \bibinfo {author} {\bibfnamefont {J.~S.}\ \bibnamefont {Loeffler}},\
  }\bibfield  {title} {\bibinfo {title} {Physics and biomedical challenges of
  cancer therapy with accelerated heavy ions},\ }\href@noop {} {\bibfield
  {journal} {\bibinfo  {journal} {Nat Rev Phys}\ }\textbf {\bibinfo {volume}
  {3(12)}},\ \bibinfo {pages} {777–790} (\bibinfo {year} {2021
  Dec})}\BibitemShut {NoStop}%
\bibitem [{\citenamefont {Grassberger}\ \emph {et~al.}(ug 1)\citenamefont
  {Grassberger}, \citenamefont {Trofimov}, \citenamefont {Lomax},\ and\
  \citenamefont {Paganetti}}]{grassberger2011}%
  \BibitemOpen
  \bibfield  {author} {\bibinfo {author} {\bibfnamefont {C.}~\bibnamefont
  {Grassberger}}, \bibinfo {author} {\bibfnamefont {A.}~\bibnamefont
  {Trofimov}}, \bibinfo {author} {\bibfnamefont {A.}~\bibnamefont {Lomax}},\
  and\ \bibinfo {author} {\bibfnamefont {H.}~\bibnamefont {Paganetti}},\
  }\bibfield  {title} {\bibinfo {title} {Variations in linear energy transfer
  within clinical proton therapy fields and the potential for biological
  treatment planning},\ }\href
  {https://doi.org/https://doi.org/10.1016/j.ijrobp.2010.10.027} {\bibfield
  {journal} {\bibinfo  {journal} {Int J Radiat Oncol Biol Phys}\ }\textbf
  {\bibinfo {volume} {80(5)}},\ \bibinfo {pages} {1559} (\bibinfo {year} {2011
  Aug 1})}\BibitemShut {NoStop}%
\bibitem [{\citenamefont {Paganetti}\ \emph {et~al.}(un 1)\citenamefont
  {Paganetti}, \citenamefont {Niemierko}, \citenamefont {Ancukiewicz},
  \citenamefont {Gerweck}, \citenamefont {Goitein}, \citenamefont {Loeffler},\
  and\ \citenamefont {Suit}}]{paganetti2002}%
  \BibitemOpen
  \bibfield  {author} {\bibinfo {author} {\bibfnamefont {H.}~\bibnamefont
  {Paganetti}}, \bibinfo {author} {\bibfnamefont {A.}~\bibnamefont
  {Niemierko}}, \bibinfo {author} {\bibfnamefont {M.}~\bibnamefont
  {Ancukiewicz}}, \bibinfo {author} {\bibfnamefont {L.~E.}\ \bibnamefont
  {Gerweck}}, \bibinfo {author} {\bibfnamefont {M.}~\bibnamefont {Goitein}},
  \bibinfo {author} {\bibfnamefont {J.}~\bibnamefont {Loeffler}},\ and\
  \bibinfo {author} {\bibfnamefont {H.}~\bibnamefont {Suit}},\ }\bibfield
  {title} {\bibinfo {title} {Relative biological effectiveness ({RBE}) values
  for proton beam therapy},\ }\href
  {https://doi.org/https://doi.org/10.1016/S0360-3016(02)02754-2} {\bibfield
  {journal} {\bibinfo  {journal} {Int J Radiat Oncol Biol Phys.}\ }\textbf
  {\bibinfo {volume} {53(2)}},\ \bibinfo {pages} {407} (\bibinfo {year} {2002
  Jun 1})}\BibitemShut {NoStop}%
\bibitem [{\citenamefont {Paganetti}\ \emph {et~al.}( Mar)\citenamefont
  {Paganetti}, \citenamefont {Blakely}, \citenamefont {Carabe-Fernandez},
  \citenamefont {Carlson}, \citenamefont {Das}, \citenamefont {Dong},
  \citenamefont {Grosshans}, \citenamefont {Held}, \citenamefont {Mohan},
  \citenamefont {Moiseenko}, \citenamefont {Niemierko}, \citenamefont
  {Stewart},\ and\ \citenamefont {Willers}}]{paganetti2019}%
  \BibitemOpen
  \bibfield  {author} {\bibinfo {author} {\bibfnamefont {H.}~\bibnamefont
  {Paganetti}}, \bibinfo {author} {\bibfnamefont {E.}~\bibnamefont {Blakely}},
  \bibinfo {author} {\bibfnamefont {A.}~\bibnamefont {Carabe-Fernandez}},
  \bibinfo {author} {\bibfnamefont {D.~J.}\ \bibnamefont {Carlson}}, \bibinfo
  {author} {\bibfnamefont {I.~J.}\ \bibnamefont {Das}}, \bibinfo {author}
  {\bibfnamefont {L.}~\bibnamefont {Dong}}, \bibinfo {author} {\bibfnamefont
  {D.}~\bibnamefont {Grosshans}}, \bibinfo {author} {\bibfnamefont {K.~D.}\
  \bibnamefont {Held}}, \bibinfo {author} {\bibfnamefont {R.}~\bibnamefont
  {Mohan}}, \bibinfo {author} {\bibfnamefont {V.}~\bibnamefont {Moiseenko}},
  \bibinfo {author} {\bibfnamefont {A.}~\bibnamefont {Niemierko}}, \bibinfo
  {author} {\bibfnamefont {R.~D.}\ \bibnamefont {Stewart}},\ and\ \bibinfo
  {author} {\bibfnamefont {H.}~\bibnamefont {Willers}},\ }\bibfield  {title}
  {\bibinfo {title} {Report of the {AAPM TG‐256} on the relative biological
  effectiveness of proton beams in radiation therapy},\ }\href
  {https://doi.org/https://doi.org/10.1002/mp.13390} {\bibfield  {journal}
  {\bibinfo  {journal} {Med Phys.}\ }\textbf {\bibinfo {volume} {46(3)}},\
  \bibinfo {pages} {e53} (\bibinfo {year} {2019 Mar})}\BibitemShut {NoStop}%
\bibitem [{\citenamefont {Paganetti}(v 21)}]{paganetti2014}%
  \BibitemOpen
  \bibfield  {author} {\bibinfo {author} {\bibfnamefont {H.}~\bibnamefont
  {Paganetti}},\ }\bibfield  {title} {\bibinfo {title} {Relative biological
  effectiveness ({RBE}) values for proton beam therapy. variations as a
  function of biological endpoint, dose, and linear energy transfer.},\
  }\href@noop {} {\bibfield  {journal} {\bibinfo  {journal} {Phys Med Biol}\
  }\textbf {\bibinfo {volume} {59(22)}},\ \bibinfo {pages} {R419} (\bibinfo
  {year} {2014 Nov 21})}\BibitemShut {NoStop}%
\bibitem [{\citenamefont {Peeler}\ \emph {et~al.}( Dec)\citenamefont {Peeler},
  \citenamefont {Mirkovic},\ and\ \citenamefont {{U. Titt \emph{et
  al}.}}}]{peeler2016}%
  \BibitemOpen
  \bibfield  {author} {\bibinfo {author} {\bibfnamefont {C.~R.}\ \bibnamefont
  {Peeler}}, \bibinfo {author} {\bibfnamefont {D.}~\bibnamefont {Mirkovic}},\
  and\ \bibinfo {author} {\bibnamefont {{U. Titt \emph{et al}.}}},\ }\bibfield
  {title} {\bibinfo {title} {Clinical evidence of variable proton biological
  effectiveness in pediatric patients treated for ependymoma.},\ }\href@noop {}
  {\bibfield  {journal} {\bibinfo  {journal} {Radiother Oncol}\ }\textbf
  {\bibinfo {volume} {121(3)}},\ \bibinfo {pages} {395} (\bibinfo {year} {2016
  Dec})}\BibitemShut {NoStop}%
\bibitem [{\citenamefont {Underwood}\ \emph {et~al.}(2017)\citenamefont
  {Underwood}, \citenamefont {Grassberger},\ and\ \citenamefont {{R. Bass
  \emph{et al}.}}}]{underwood2017}%
  \BibitemOpen
  \bibfield  {author} {\bibinfo {author} {\bibfnamefont {T.}~\bibnamefont
  {Underwood}}, \bibinfo {author} {\bibfnamefont {C.}~\bibnamefont
  {Grassberger}},\ and\ \bibinfo {author} {\bibnamefont {{R. Bass \emph{et
  al}.}}},\ }\bibfield  {title} {\bibinfo {title} {{OC-0245}: Clinical evidence
  that end-of-range proton {RBE} exceeds 1.1: lung density changes following
  chest {RT}.},\ }\href@noop {} {\bibfield  {journal} {\bibinfo  {journal}
  {Radiotherapy and Oncology}\ }\textbf {\bibinfo {volume} {Supplement, 123}},\
  \bibinfo {pages} {S123} (\bibinfo {year} {2017})}\BibitemShut {NoStop}%
\bibitem [{\citenamefont {Bragg}\ and\ \citenamefont
  {Kleeman}(1905)}]{bragg1905}%
  \BibitemOpen
  \bibfield  {author} {\bibinfo {author} {\bibfnamefont {W.}~\bibnamefont
  {Bragg}}\ and\ \bibinfo {author} {\bibfnamefont {R.}~\bibnamefont
  {Kleeman}},\ }\bibfield  {title} {\bibinfo {title} {On the $\alpha$ particles
  of radium, and their loss of range in passing through various atoms and
  molecules},\ }\href@noop {} {\bibfield  {journal} {\bibinfo  {journal} {Phil.
  Mag}\ }\textbf {\bibinfo {volume} {10}},\ \bibinfo {pages} {318} (\bibinfo
  {year} {1905})}\BibitemShut {NoStop}%
\bibitem [{\citenamefont {Wilson}(1946)}]{william1946}%
  \BibitemOpen
  \bibfield  {author} {\bibinfo {author} {\bibfnamefont {R.~R.}\ \bibnamefont
  {Wilson}},\ }\bibfield  {title} {\bibinfo {title} {Radiological {Use} of
  {Fast} {Protons}},\ }\href@noop {} {\bibfield  {journal} {\bibinfo  {journal}
  {Radiology}\ }\textbf {\bibinfo {volume} {47}},\ \bibinfo {pages} {487–91}
  (\bibinfo {year} {1946})}\BibitemShut {NoStop}%
\bibitem [{\citenamefont {Kumakhov}\ and\ \citenamefont
  {Komarov}(1981)}]{bragg:gordon}%
  \BibitemOpen
  \bibfield  {author} {\bibinfo {author} {\bibfnamefont {M.}~\bibnamefont
  {Kumakhov}}\ and\ \bibinfo {author} {\bibfnamefont {F.}~\bibnamefont
  {Komarov}},\ }\href@noop {} {\emph {\bibinfo {title} {Energy Loss and Ion
  Ranges in Solids}}},\ \bibinfo {edition} {revised english edition, translated
  from the russian by ten teague}\ ed.\ (\bibinfo  {publisher} {Gordon and
  Breach Science Publishers},\ \bibinfo {address} {New York, London, Paris},\
  \bibinfo {year} {1981})\BibitemShut {NoStop}%
\bibitem [{\citenamefont {Brown}\ and\ \citenamefont {Suit}( Dec)}]{brown2004}%
  \BibitemOpen
  \bibfield  {author} {\bibinfo {author} {\bibfnamefont {A.}~\bibnamefont
  {Brown}}\ and\ \bibinfo {author} {\bibfnamefont {H.}~\bibnamefont {Suit}},\
  }\bibfield  {title} {\bibinfo {title} {The centenary of the discovery of the
  {Bragg} peak},\ }\href@noop {} {\bibfield  {journal} {\bibinfo  {journal}
  {Radiotherapy and Oncology}\ }\textbf {\bibinfo {volume} {73(3)}},\ \bibinfo
  {pages} {265–268} (\bibinfo {year} {2004 Dec})}\BibitemShut {NoStop}%
\bibitem [{\citenamefont {Hall}(ay 1)}]{hall2006}%
  \BibitemOpen
  \bibfield  {author} {\bibinfo {author} {\bibfnamefont {E.~J.}\ \bibnamefont
  {Hall}},\ }\bibfield  {title} {\bibinfo {title} {Intensity-modulated
  radiation therapy, protons, and the risk of second cancers},\ }\href
  {https://doi.org/https://doi.org/10.1016/j.ijrobp.2006.01.027} {\bibfield
  {journal} {\bibinfo  {journal} {Int J Radiat Oncol Biol Phys.}\ }\textbf
  {\bibinfo {volume} {65(1)}},\ \bibinfo {pages} {1} (\bibinfo {year} {2006 May
  1})}\BibitemShut {NoStop}%
\bibitem [{\citenamefont {Merchant}( Aug)}]{merchant2009}%
  \BibitemOpen
  \bibfield  {author} {\bibinfo {author} {\bibfnamefont {T.~E.}\ \bibnamefont
  {Merchant}},\ }\bibfield  {title} {\bibinfo {title} {Proton beam therapy in
  pediatric oncology},\ }\href
  {https://doi.org/https://journals.lww.com/journalppo/abstract/2009/08000/proton_beam_therapy_in_pediatric_oncology.6.aspx}
  {\bibfield  {journal} {\bibinfo  {journal} {Cancer J.}\ }\textbf {\bibinfo
  {volume} {15(4)}},\ \bibinfo {pages} {298} (\bibinfo {year} {2009
  Jul-Aug})}\BibitemShut {NoStop}%
\bibitem [{\citenamefont {Zhang}\ \emph {et~al.}( Oct)\citenamefont {Zhang},
  \citenamefont {Howell}, \citenamefont {Taddei}, \citenamefont {Giebeler},
  \citenamefont {Mahajan},\ and\ \citenamefont {Newhauser}}]{zhang2014}%
  \BibitemOpen
  \bibfield  {author} {\bibinfo {author} {\bibfnamefont {R.}~\bibnamefont
  {Zhang}}, \bibinfo {author} {\bibfnamefont {R.~M.}\ \bibnamefont {Howell}},
  \bibinfo {author} {\bibfnamefont {P.~J.}\ \bibnamefont {Taddei}}, \bibinfo
  {author} {\bibfnamefont {A.}~\bibnamefont {Giebeler}}, \bibinfo {author}
  {\bibfnamefont {A.}~\bibnamefont {Mahajan}},\ and\ \bibinfo {author}
  {\bibfnamefont {W.~D.}\ \bibnamefont {Newhauser}},\ }\bibfield  {title}
  {\bibinfo {title} {A comparative study on the risks of radiogenic second
  cancers and cardiac mortality in a set of pediatric medulloblastoma patients
  treated with photon or proton craniospinal irradiation},\ }\href@noop {}
  {\bibfield  {journal} {\bibinfo  {journal} {Radiother Oncol.}\ }\textbf
  {\bibinfo {volume} {113(1)}},\ \bibinfo {pages} {84} (\bibinfo {year} {2014
  Oct})}\BibitemShut {NoStop}%
\bibitem [{\citenamefont {{C-M Charlie Ma}}\ and\ \citenamefont {{Tony
  Lomax}}(2012)}]{ma2012}%
  \BibitemOpen
  \bibfield  {author} {\bibinfo {author} {\bibnamefont {{C-M Charlie Ma}}}\
  and\ \bibinfo {author} {\bibnamefont {{Tony Lomax}}},\ }\href@noop {} {\emph
  {\bibinfo {title} {Proton and Carbon Ion Therapy}}},\ \bibinfo {edition}
  {1st}\ ed.\ (\bibinfo  {publisher} {CRC Press},\ \bibinfo {year} {October 9,
  2012})\BibitemShut {NoStop}%
\bibitem [{\citenamefont {Paganetti}(un 7)}]{paganetti2012}%
  \BibitemOpen
  \bibfield  {author} {\bibinfo {author} {\bibfnamefont {H.}~\bibnamefont
  {Paganetti}},\ }\bibfield  {title} {\bibinfo {title} {Range uncertainties in
  proton therapy and the role of monte carlo simulations},\ }\href
  {https://doi.org/https://iopscience.iop.org/article/10.1088/0031-9155/57/11/R99}
  {\bibfield  {journal} {\bibinfo  {journal} {Phys Med Biol.}\ }\textbf
  {\bibinfo {volume} {57(11)}},\ \bibinfo {pages} {R99} (\bibinfo {year} {2012
  Jun 7})}\BibitemShut {NoStop}%
\bibitem [{\citenamefont {Brunetto}\ and\ \citenamefont
  {Strazzulla}(2005)}]{Brunetto}%
  \BibitemOpen
  \bibfield  {author} {\bibinfo {author} {\bibfnamefont {R.}~\bibnamefont
  {Brunetto}}\ and\ \bibinfo {author} {\bibfnamefont {G.}~\bibnamefont
  {Strazzulla}},\ }\bibfield  {title} {\bibinfo {title} {Elastic collisions in
  ion irradiation experiments: A mechanism for space weathering of silicates},\
  }\href@noop {} {\bibfield  {journal} {\bibinfo  {journal} {Icarus}\ }\textbf
  {\bibinfo {volume} {179}},\ \bibinfo {pages} {265} (\bibinfo {year}
  {2005})}\BibitemShut {NoStop}%
\bibitem [{\citenamefont {Kim}\ \emph {et~al.}(y 27)\citenamefont {Kim},
  \citenamefont {Lee},\ and\ \citenamefont {Hong}}]{Kim}%
  \BibitemOpen
  \bibfield  {author} {\bibinfo {author} {\bibfnamefont {S.}~\bibnamefont
  {Kim}}, \bibinfo {author} {\bibfnamefont {S.}~\bibnamefont {Lee}},\ and\
  \bibinfo {author} {\bibfnamefont {J.}~\bibnamefont {Hong}},\ }\bibfield
  {title} {\bibinfo {title} {An array of ferromagnetic nanoislands
  nondestructively patterned via a local phase transformation by low-energy
  proton irradiation},\ }\href@noop {} {\bibfield  {journal} {\bibinfo
  {journal} {ACS Nano}\ }\textbf {\bibinfo {volume} {8(5)}},\ \bibinfo {pages}
  {4698} (\bibinfo {year} {2014 May 27})}\BibitemShut {NoStop}%
\bibitem [{\citenamefont {Papaléo}\ \emph {et~al.}(2015)\citenamefont
  {Papaléo}, \citenamefont {Thomaz}, \citenamefont {Gutierres}, \citenamefont
  {de~Menezes}, \citenamefont {Severin}, \citenamefont {Trautmann},
  \citenamefont {Tramontina}, \citenamefont {Bringa},\ and\ \citenamefont
  {Grande}}]{Papale:2015}%
  \BibitemOpen
  \bibfield  {author} {\bibinfo {author} {\bibfnamefont {R.~M.}\ \bibnamefont
  {Papaléo}}, \bibinfo {author} {\bibfnamefont {R.}~\bibnamefont {Thomaz}},
  \bibinfo {author} {\bibfnamefont {L.~I.}\ \bibnamefont {Gutierres}}, \bibinfo
  {author} {\bibfnamefont {V.~M.}\ \bibnamefont {de~Menezes}}, \bibinfo
  {author} {\bibfnamefont {D.}~\bibnamefont {Severin}}, \bibinfo {author}
  {\bibfnamefont {C.}~\bibnamefont {Trautmann}}, \bibinfo {author}
  {\bibfnamefont {D.}~\bibnamefont {Tramontina}}, \bibinfo {author}
  {\bibfnamefont {E.~M.}\ \bibnamefont {Bringa}},\ and\ \bibinfo {author}
  {\bibfnamefont {P.~L.}\ \bibnamefont {Grande}},\ }\bibfield  {title}
  {\bibinfo {title} {Confinement effects of ion tracks in ultrathin polymer
  films},\ }\href@noop {} {\bibfield  {journal} {\bibinfo  {journal} {Phys.
  Rev. Lett.}\ }\textbf {\bibinfo {volume} {114}},\ \bibinfo {pages} {118302}
  (\bibinfo {year} {2015})}\BibitemShut {NoStop}%
\bibitem [{\citenamefont {Meisl}\ \emph {et~al.}(2014)\citenamefont {Meisl},
  \citenamefont {Schmid}, \citenamefont {Encke}, \citenamefont {Höschen},
  \citenamefont {Gao},\ and\ \citenamefont {Linsmeier}}]{Meisl}%
  \BibitemOpen
  \bibfield  {author} {\bibinfo {author} {\bibfnamefont {G.}~\bibnamefont
  {Meisl}}, \bibinfo {author} {\bibfnamefont {K.}~\bibnamefont {Schmid}},
  \bibinfo {author} {\bibfnamefont {O.}~\bibnamefont {Encke}}, \bibinfo
  {author} {\bibfnamefont {T.}~\bibnamefont {Höschen}}, \bibinfo {author}
  {\bibfnamefont {L.}~\bibnamefont {Gao}},\ and\ \bibinfo {author}
  {\bibfnamefont {C.}~\bibnamefont {Linsmeier}},\ }\bibfield  {title} {\bibinfo
  {title} {Implantation and erosion of nitrogen in tungsten},\ }\href@noop {}
  {\bibfield  {journal} {\bibinfo  {journal} {New J. Phys.}\ }\textbf {\bibinfo
  {volume} {16}},\ \bibinfo {pages} {093018} (\bibinfo {year}
  {2014})}\BibitemShut {NoStop}%
\bibitem [{\citenamefont {{Mayer, M. \emph{et al}.}}(2020)}]{mayer_ion_2019}%
  \BibitemOpen
  \bibfield  {author} {\bibinfo {author} {\bibnamefont {{Mayer, M. \emph{et
  al}.}}},\ }\bibfield  {title} {\bibinfo {title} {Ion beam analysis of fusion
  plasma-facing materials and components: facilities and research challenges},\
  }\href {https://doi.org/10.1088/1741-4326/ab5817} {\bibfield  {journal}
  {\bibinfo  {journal} {Nucl. Fusion}\ }\textbf {\bibinfo {volume} {60}},\
  \bibinfo {pages} {025001} (\bibinfo {year} {2020})}\BibitemShut {NoStop}%
\bibitem [{\citenamefont {Newhauser}\ and\ \citenamefont {Zhang}(r
  21)}]{Newhauser:2015}%
  \BibitemOpen
  \bibfield  {author} {\bibinfo {author} {\bibfnamefont {W.~D.}\ \bibnamefont
  {Newhauser}}\ and\ \bibinfo {author} {\bibfnamefont {R.}~\bibnamefont
  {Zhang}},\ }\bibfield  {title} {\bibinfo {title} {The physics of proton
  therapy},\ }\href@noop {} {\bibfield  {journal} {\bibinfo  {journal} {Phys.
  Med. Biol.}\ }\textbf {\bibinfo {volume} {60}},\ \bibinfo {pages} {R155}
  (\bibinfo {year} {2015 Apr 21})}\BibitemShut {NoStop}%
\bibitem [{\citenamefont {Durante}\ and\ \citenamefont {Loeffler}(
  Jan)}]{Durante}%
  \BibitemOpen
  \bibfield  {author} {\bibinfo {author} {\bibfnamefont {M.}~\bibnamefont
  {Durante}}\ and\ \bibinfo {author} {\bibfnamefont {J.~S.}\ \bibnamefont
  {Loeffler}},\ }\bibfield  {title} {\bibinfo {title} {Charged particles in
  radiation oncology},\ }\href@noop {} {\bibfield  {journal} {\bibinfo
  {journal} {Nat. Rev. Clin. Oncol.}\ }\textbf {\bibinfo {volume} {7(1)}},\
  \bibinfo {pages} {37} (\bibinfo {year} {2010 Jan})}\BibitemShut {NoStop}%
\bibitem [{\citenamefont {Pruneda}\ \emph {et~al.}(2007)\citenamefont
  {Pruneda}, \citenamefont {Sánchez-Portal}, \citenamefont {Arnau},
  \citenamefont {Juaristi},\ and\ \citenamefont {Artacho}}]{pruneda2007}%
  \BibitemOpen
  \bibfield  {author} {\bibinfo {author} {\bibfnamefont {J.~M.}\ \bibnamefont
  {Pruneda}}, \bibinfo {author} {\bibfnamefont {D.}~\bibnamefont
  {Sánchez-Portal}}, \bibinfo {author} {\bibfnamefont {A.}~\bibnamefont
  {Arnau}}, \bibinfo {author} {\bibfnamefont {J.~I.}\ \bibnamefont
  {Juaristi}},\ and\ \bibinfo {author} {\bibfnamefont {E.}~\bibnamefont
  {Artacho}},\ }\bibfield  {title} {\bibinfo {title} {Electronic stopping power
  in lif from first principles},\ }\href@noop {} {\bibfield  {journal}
  {\bibinfo  {journal} {Phys. Rev. Lett.}\ }\textbf {\bibinfo {volume} {99}},\
  \bibinfo {pages} {235501} (\bibinfo {year} {2007})}\BibitemShut {NoStop}%
\bibitem [{\citenamefont {Quijada}\ \emph {et~al.}(2007)\citenamefont
  {Quijada}, \citenamefont {Borisov}, \citenamefont {Nagy}, \citenamefont
  {Muiño},\ and\ \citenamefont {Echenique}}]{quijada2007}%
  \BibitemOpen
  \bibfield  {author} {\bibinfo {author} {\bibfnamefont {M.}~\bibnamefont
  {Quijada}}, \bibinfo {author} {\bibfnamefont {A.~G.}\ \bibnamefont
  {Borisov}}, \bibinfo {author} {\bibfnamefont {I.}~\bibnamefont {Nagy}},
  \bibinfo {author} {\bibfnamefont {R.~D.}\ \bibnamefont {Muiño}},\ and\
  \bibinfo {author} {\bibfnamefont {P.~M.}\ \bibnamefont {Echenique}},\
  }\bibfield  {title} {\bibinfo {title} {Time-dependent density-functional
  calculation of the stopping power for protons and antiprotons in metals},\
  }\href@noop {} {\bibfield  {journal} {\bibinfo  {journal} {Phys. Rev. A}\
  }\textbf {\bibinfo {volume} {75}},\ \bibinfo {pages} {042902} (\bibinfo
  {year} {2007})}\BibitemShut {NoStop}%
\bibitem [{\citenamefont {Goebl}\ \emph {et~al.}(2014)\citenamefont {Goebl},
  \citenamefont {Roessler}, \citenamefont {Roth},\ and\ \citenamefont
  {Bauer}}]{goebl2014}%
  \BibitemOpen
  \bibfield  {author} {\bibinfo {author} {\bibfnamefont {D.}~\bibnamefont
  {Goebl}}, \bibinfo {author} {\bibfnamefont {W.}~\bibnamefont {Roessler}},
  \bibinfo {author} {\bibfnamefont {D.}~\bibnamefont {Roth}},\ and\ \bibinfo
  {author} {\bibfnamefont {P.}~\bibnamefont {Bauer}},\ }\bibfield  {title}
  {\bibinfo {title} {Influence of the excitation threshold of d electrons on
  electronic stopping of slow light ions},\ }\href@noop {} {\bibfield
  {journal} {\bibinfo  {journal} {Phys. Rev. A}\ }\textbf {\bibinfo {volume}
  {90}},\ \bibinfo {pages} {042706} (\bibinfo {year} {2014})}\BibitemShut
  {NoStop}%
\bibitem [{\citenamefont {Roth}\ \emph
  {et~al.}(2017{\natexlab{a}})\citenamefont {Roth}, \citenamefont {Bruckner},
  \citenamefont {Moro}, \citenamefont {Gruber}, \citenamefont {Goebl},
  \citenamefont {Juaristi}, \citenamefont {Alducin}, \citenamefont
  {Steinberger}, \citenamefont {Duchoslav}, \citenamefont {Primetzhofer}, ,\
  and\ \citenamefont {Bauer}}]{roth2017a}%
  \BibitemOpen
  \bibfield  {author} {\bibinfo {author} {\bibfnamefont {D.}~\bibnamefont
  {Roth}}, \bibinfo {author} {\bibfnamefont {B.}~\bibnamefont {Bruckner}},
  \bibinfo {author} {\bibfnamefont {M.~V.}\ \bibnamefont {Moro}}, \bibinfo
  {author} {\bibfnamefont {S.}~\bibnamefont {Gruber}}, \bibinfo {author}
  {\bibfnamefont {D.}~\bibnamefont {Goebl}}, \bibinfo {author} {\bibfnamefont
  {J.~I.}\ \bibnamefont {Juaristi}}, \bibinfo {author} {\bibfnamefont
  {M.}~\bibnamefont {Alducin}}, \bibinfo {author} {\bibfnamefont
  {R.}~\bibnamefont {Steinberger}}, \bibinfo {author} {\bibfnamefont
  {J.}~\bibnamefont {Duchoslav}}, \bibinfo {author} {\bibfnamefont
  {D.}~\bibnamefont {Primetzhofer}}, ,\ and\ \bibinfo {author} {\bibfnamefont
  {P.}~\bibnamefont {Bauer}},\ }\bibfield  {title} {\bibinfo {title}
  {Electronic stopping of slow protons in transition and rare earth metals:
  Breakdown of the free electron gas concept},\ }\href@noop {} {\bibfield
  {journal} {\bibinfo  {journal} {Phys. Rev. Lett.}\ }\textbf {\bibinfo
  {volume} {118}},\ \bibinfo {pages} {103401} (\bibinfo {year}
  {2017}{\natexlab{a}})}\BibitemShut {NoStop}%
\bibitem [{\citenamefont {Roth}\ \emph
  {et~al.}(2017{\natexlab{b}})\citenamefont {Roth}, \citenamefont {Bruckner},
  \citenamefont {Undeutsch}, \citenamefont {Paneta}, \citenamefont {Mardare},
  \citenamefont {McGahan}, \citenamefont {Dosmailov}, \citenamefont {Juaristi},
  \citenamefont {Alducin}, \citenamefont {Pedarnig}, \citenamefont {Haglund},
  \citenamefont {Primetzhofer},\ and\ \citenamefont {Bauer}}]{roth2017b}%
  \BibitemOpen
  \bibfield  {author} {\bibinfo {author} {\bibfnamefont {D.}~\bibnamefont
  {Roth}}, \bibinfo {author} {\bibfnamefont {B.}~\bibnamefont {Bruckner}},
  \bibinfo {author} {\bibfnamefont {G.}~\bibnamefont {Undeutsch}}, \bibinfo
  {author} {\bibfnamefont {V.}~\bibnamefont {Paneta}}, \bibinfo {author}
  {\bibfnamefont {A.~I.}\ \bibnamefont {Mardare}}, \bibinfo {author}
  {\bibfnamefont {C.~L.}\ \bibnamefont {McGahan}}, \bibinfo {author}
  {\bibfnamefont {M.}~\bibnamefont {Dosmailov}}, \bibinfo {author}
  {\bibfnamefont {J.~I.}\ \bibnamefont {Juaristi}}, \bibinfo {author}
  {\bibfnamefont {M.}~\bibnamefont {Alducin}}, \bibinfo {author} {\bibfnamefont
  {J.~D.}\ \bibnamefont {Pedarnig}}, \bibinfo {author} {\bibfnamefont {R.~F.}\
  \bibnamefont {Haglund}}, \bibinfo {author} {\bibfnamefont {D.}~\bibnamefont
  {Primetzhofer}},\ and\ \bibinfo {author} {\bibfnamefont {P.}~\bibnamefont
  {Bauer}},\ }\bibfield  {title} {\bibinfo {title} {Electronic stopping of slow
  protons in oxides: Scaling properties},\ }\href@noop {} {\bibfield  {journal}
  {\bibinfo  {journal} {Phys. Rev. Lett.}\ }\textbf {\bibinfo {volume} {119}},\
  \bibinfo {pages} {163401} (\bibinfo {year} {2017}{\natexlab{b}})}\BibitemShut
  {NoStop}%
\bibitem [{\citenamefont {Sortica}\ \emph {et~al.}(2017)\citenamefont
  {Sortica}, \citenamefont {Paneta}, \citenamefont {Bruckner}, \citenamefont
  {Lohmann}, \citenamefont {Hans}, \citenamefont {Nyberg}, \citenamefont
  {Bauer},\ and\ \citenamefont {Primetzhofer}}]{sortica2017}%
  \BibitemOpen
  \bibfield  {author} {\bibinfo {author} {\bibfnamefont {M.~A.}\ \bibnamefont
  {Sortica}}, \bibinfo {author} {\bibfnamefont {V.}~\bibnamefont {Paneta}},
  \bibinfo {author} {\bibfnamefont {B.}~\bibnamefont {Bruckner}}, \bibinfo
  {author} {\bibfnamefont {S.}~\bibnamefont {Lohmann}}, \bibinfo {author}
  {\bibfnamefont {M.}~\bibnamefont {Hans}}, \bibinfo {author} {\bibfnamefont
  {T.}~\bibnamefont {Nyberg}}, \bibinfo {author} {\bibfnamefont
  {P.}~\bibnamefont {Bauer}},\ and\ \bibinfo {author} {\bibfnamefont
  {D.}~\bibnamefont {Primetzhofer}},\ }\bibfield  {title} {\bibinfo {title}
  {Electronic energy-loss mechanisms for {H}, {He}, and {Ne} in {TiN}},\
  }\href@noop {} {\bibfield  {journal} {\bibinfo  {journal} {Phys. Rev. A}\
  }\textbf {\bibinfo {volume} {96}},\ \bibinfo {pages} {032703} (\bibinfo
  {year} {2017})}\BibitemShut {NoStop}%
\bibitem [{\citenamefont {Matias}\ \emph {et~al.}(2017)\citenamefont {Matias},
  \citenamefont {Fadanelli}, \citenamefont {Grande}, \citenamefont {Koval},
  \citenamefont {Muiño}, \citenamefont {Borisov}, \citenamefont {Arista},\
  and\ \citenamefont {Schiwietz}}]{matias2017}%
  \BibitemOpen
  \bibfield  {author} {\bibinfo {author} {\bibfnamefont {F.}~\bibnamefont
  {Matias}}, \bibinfo {author} {\bibfnamefont {R.~C.}\ \bibnamefont
  {Fadanelli}}, \bibinfo {author} {\bibfnamefont {P.~L.}\ \bibnamefont
  {Grande}}, \bibinfo {author} {\bibfnamefont {N.~E.}\ \bibnamefont {Koval}},
  \bibinfo {author} {\bibfnamefont {R.~D.}\ \bibnamefont {Muiño}}, \bibinfo
  {author} {\bibfnamefont {A.~G.}\ \bibnamefont {Borisov}}, \bibinfo {author}
  {\bibfnamefont {N.~R.}\ \bibnamefont {Arista}},\ and\ \bibinfo {author}
  {\bibfnamefont {G.}~\bibnamefont {Schiwietz}},\ }\bibfield  {title} {\bibinfo
  {title} {Ground- and excited-state scattering potentials for the stopping of
  protons in an electron gas},\ }\href@noop {} {\bibfield  {journal} {\bibinfo
  {journal} {J. Phys. B: At. Mol. Opt. Phys.}\ }\textbf {\bibinfo {volume}
  {50}},\ \bibinfo {pages} {185201} (\bibinfo {year} {2017})}\BibitemShut
  {NoStop}%
\bibitem [{\citenamefont {Matias}\ \emph {et~al.}(2019)\citenamefont {Matias},
  \citenamefont {Grande}, \citenamefont {Vos}, \citenamefont {Koval},\ and\
  \citenamefont {Arista}}]{Matias:2019}%
  \BibitemOpen
  \bibfield  {author} {\bibinfo {author} {\bibfnamefont {F.}~\bibnamefont
  {Matias}}, \bibinfo {author} {\bibfnamefont {P.~L.}\ \bibnamefont {Grande}},
  \bibinfo {author} {\bibfnamefont {P.}~\bibnamefont {Vos}, \bibfnamefont
  {M~Koval}}, \bibinfo {author} {\bibfnamefont {N.~E.}\ \bibnamefont {Koval}},\
  and\ \bibinfo {author} {\bibfnamefont {N.~R.}\ \bibnamefont {Arista}},\
  }\bibfield  {title} {\bibinfo {title} {Nonlinear stopping effects of slow
  ions in a no-free-electron system: Titanium nitride},\ }\href@noop {}
  {\bibfield  {journal} {\bibinfo  {journal} {Physical Review A}\ }\textbf
  {\bibinfo {volume} {100}},\ \bibinfo {pages} {030701(R)} (\bibinfo {year}
  {2019})}\BibitemShut {NoStop}%
\bibitem [{\citenamefont {Ullah}\ \emph {et~al.}(2018)\citenamefont {Ullah},
  \citenamefont {Artacho},\ and\ \citenamefont {Correa}}]{ullah2018}%
  \BibitemOpen
  \bibfield  {author} {\bibinfo {author} {\bibfnamefont {R.}~\bibnamefont
  {Ullah}}, \bibinfo {author} {\bibfnamefont {E.}~\bibnamefont {Artacho}},\
  and\ \bibinfo {author} {\bibfnamefont {A.~A.}\ \bibnamefont {Correa}},\
  }\bibfield  {title} {\bibinfo {title} {Core electrons in the electronic
  stopping of heavy ions},\ }\href@noop {} {\bibfield  {journal} {\bibinfo
  {journal} {Phys. Rev. Lett.}\ }\textbf {\bibinfo {volume} {121}},\ \bibinfo
  {pages} {116401} (\bibinfo {year} {2018})}\BibitemShut {NoStop}%
\bibitem [{\citenamefont {Koval}\ \emph {et~al.}(2013)\citenamefont {Koval},
  \citenamefont {Sánchez-Portal}, \citenamefont {Borisov},\ and\ \citenamefont
  {Muiño}}]{koval2013}%
  \BibitemOpen
  \bibfield  {author} {\bibinfo {author} {\bibfnamefont {N.~E.}\ \bibnamefont
  {Koval}}, \bibinfo {author} {\bibfnamefont {D.}~\bibnamefont
  {Sánchez-Portal}}, \bibinfo {author} {\bibfnamefont {A.~G.}\ \bibnamefont
  {Borisov}},\ and\ \bibinfo {author} {\bibfnamefont {R.~D.}\ \bibnamefont
  {Muiño}},\ }\bibfield  {title} {\bibinfo {title} {Dynamic screening and
  energy loss of antiprotons colliding with excited ai clurters},\ }\href@noop
  {} {\bibfield  {journal} {\bibinfo  {journal} {Nucl. Instrum. Methods Phys.
  Res. B}\ }\textbf {\bibinfo {volume} {317}},\ \bibinfo {pages} {56} (\bibinfo
  {year} {2013})}\BibitemShut {NoStop}%
\bibitem [{\citenamefont {Koval}\ \emph {et~al.}(2017)\citenamefont {Koval},
  \citenamefont {Borisov}, \citenamefont {Rosa}, \citenamefont {Stori},
  \citenamefont {Dias}, \citenamefont {Grande}, \citenamefont
  {Sánchez-Portal},\ and\ \citenamefont {Muiño}}]{Koval2017}%
  \BibitemOpen
  \bibfield  {author} {\bibinfo {author} {\bibfnamefont {N.~E.}\ \bibnamefont
  {Koval}}, \bibinfo {author} {\bibfnamefont {A.~G.}\ \bibnamefont {Borisov}},
  \bibinfo {author} {\bibfnamefont {L.~F.~S.}\ \bibnamefont {Rosa}}, \bibinfo
  {author} {\bibfnamefont {E.~M.}\ \bibnamefont {Stori}}, \bibinfo {author}
  {\bibfnamefont {J.~F.}\ \bibnamefont {Dias}}, \bibinfo {author}
  {\bibfnamefont {P.~L.}\ \bibnamefont {Grande}}, \bibinfo {author}
  {\bibfnamefont {D.}~\bibnamefont {Sánchez-Portal}},\ and\ \bibinfo {author}
  {\bibfnamefont {R.~D.}\ \bibnamefont {Muiño}},\ }\bibfield  {title}
  {\bibinfo {title} {Vicinage effect in the energy loss of h2 dimers:
  Experiment and calculations based on time-dependent density-functional
  theory},\ }\href@noop {} {\bibfield  {journal} {\bibinfo  {journal} {Phys.
  Rev. A}\ }\textbf {\bibinfo {volume} {95}},\ \bibinfo {pages} {062707}
  (\bibinfo {year} {2017})}\BibitemShut {NoStop}%
\bibitem [{\citenamefont {Vos}\ and\ \citenamefont
  {Grande}(2019)}]{Maarten:2019b}%
  \BibitemOpen
  \bibfield  {author} {\bibinfo {author} {\bibfnamefont {M.}~\bibnamefont
  {Vos}}\ and\ \bibinfo {author} {\bibfnamefont {P.~L.}\ \bibnamefont
  {Grande}},\ }\bibfield  {title} {\bibinfo {title} {Extension schemes of the
  dielectric function, and their implications for ion stopping calculations},\
  }\href@noop {} {\bibfield  {journal} {\bibinfo  {journal} {Journal of Physics
  and Chemistry of Solids}\ }\textbf {\bibinfo {volume} {133}},\ \bibinfo
  {pages} {187} (\bibinfo {year} {2019})}\BibitemShut {NoStop}%
\bibitem [{\citenamefont {Penn}(1987)}]{penn1987}%
  \BibitemOpen
  \bibfield  {author} {\bibinfo {author} {\bibfnamefont {D.~R.}\ \bibnamefont
  {Penn}},\ }\bibfield  {title} {\bibinfo {title} {Electron mean-free-path
  calculations using a model dielectric function},\ }\href@noop {} {\bibfield
  {journal} {\bibinfo  {journal} {Phys. Rev. B}\ }\textbf {\bibinfo {volume}
  {35(2)}},\ \bibinfo {pages} {482} (\bibinfo {year} {1987})}\BibitemShut
  {NoStop}%
\bibitem [{\citenamefont {Shinotsuka}\ \emph {et~al.}(2012)\citenamefont
  {Shinotsuka}, \citenamefont {Tanuma}, \citenamefont {Powell},\ and\
  \citenamefont {Penn}}]{SHINOTSUKA201275}%
  \BibitemOpen
  \bibfield  {author} {\bibinfo {author} {\bibfnamefont {H.}~\bibnamefont
  {Shinotsuka}}, \bibinfo {author} {\bibfnamefont {S.}~\bibnamefont {Tanuma}},
  \bibinfo {author} {\bibfnamefont {C.}~\bibnamefont {Powell}},\ and\ \bibinfo
  {author} {\bibfnamefont {D.}~\bibnamefont {Penn}},\ }\bibfield  {title}
  {\bibinfo {title} {Calculations of electron stopping powers for 41 elemental
  solids over the {50 eV} to {30 keV} range with the full {Penn} algorithm},\
  }\href {https://doi.org/https://doi.org/10.1016/j.nimb.2011.09.016}
  {\bibfield  {journal} {\bibinfo  {journal} {Nuclear Instruments and Methods
  in Physics Research Section B: Beam Interactions with Materials and Atoms}\
  }\textbf {\bibinfo {volume} {270}},\ \bibinfo {pages} {75} (\bibinfo {year}
  {2012})}\BibitemShut {NoStop}%
\bibitem [{\citenamefont {Sun}\ \emph {et~al.}(html)\citenamefont {Sun},
  \citenamefont {Xu}, \citenamefont {Da}, \citenamefont {Mao},\ and\
  \citenamefont {Ding}}]{ding:ELF}%
  \BibitemOpen
  \bibfield  {author} {\bibinfo {author} {\bibfnamefont {Y.}~\bibnamefont
  {Sun}}, \bibinfo {author} {\bibfnamefont {H.}~\bibnamefont {Xu}}, \bibinfo
  {author} {\bibfnamefont {B.}~\bibnamefont {Da}}, \bibinfo {author}
  {\bibfnamefont {S.}~\bibnamefont {Mao}},\ and\ \bibinfo {author}
  {\bibfnamefont {Z.}~\bibnamefont {Ding}},\ }\bibfield  {title} {\bibinfo
  {title} {Database of fitted energy-loss function for 26 materials},\
  }\href@noop {} {\  (\bibinfo {year}
  {http://micro.ustc.edu.cn/ELF/ELF.html})}\BibitemShut {NoStop}%
\bibitem [{\citenamefont {Lodhi}\ and\ \citenamefont
  {Powers}(1974)}]{lodhi1974}%
  \BibitemOpen
  \bibfield  {author} {\bibinfo {author} {\bibfnamefont {A.~S.}\ \bibnamefont
  {Lodhi}}\ and\ \bibinfo {author} {\bibfnamefont {D.}~\bibnamefont {Powers}},\
  }\bibfield  {title} {\bibinfo {title} {Energy loss of $\alpha$ particles in
  gaseous {C-H} and {C-H-F} compounds},\ }\href@noop {} {\bibfield  {journal}
  {\bibinfo  {journal} {Phys. Rev.}\ }\textbf {\bibinfo {volume} {A10}},\
  \bibinfo {pages} {2131} (\bibinfo {year} {1974})}\BibitemShut {NoStop}%
\bibitem [{\citenamefont {Kirby}\ \emph {et~al.}(n 21)\citenamefont {Kirby},
  \citenamefont {Green}, \citenamefont {Palmans}, \citenamefont {Hugtenburg},
  \citenamefont {Wojnecki},\ and\ \citenamefont {Parker}}]{kirby2010}%
  \BibitemOpen
  \bibfield  {author} {\bibinfo {author} {\bibfnamefont {D.}~\bibnamefont
  {Kirby}}, \bibinfo {author} {\bibfnamefont {S.}~\bibnamefont {Green}},
  \bibinfo {author} {\bibfnamefont {H.}~\bibnamefont {Palmans}}, \bibinfo
  {author} {\bibfnamefont {R.}~\bibnamefont {Hugtenburg}}, \bibinfo {author}
  {\bibfnamefont {C.}~\bibnamefont {Wojnecki}},\ and\ \bibinfo {author}
  {\bibfnamefont {D.}~\bibnamefont {Parker}},\ }\bibfield  {title} {\bibinfo
  {title} {{LET dependence of GafChromic films and an ion chamber in low-energy
  proton dosimetry}},\ }\href@noop {} {\bibfield  {journal} {\bibinfo
  {journal} {Phys Med Biol.}\ }\textbf {\bibinfo {volume} {55(2)}},\ \bibinfo
  {pages} {417} (\bibinfo {year} {2010 Jan 21})}\BibitemShut {NoStop}%
\bibitem [{\citenamefont {{P. de Vera \emph{et al}.}}(2013)}]{deVera2013}%
  \BibitemOpen
  \bibfield  {author} {\bibinfo {author} {\bibnamefont {{P. de Vera \emph{et
  al}.}}},\ }\bibfield  {title} {\bibinfo {title} {Water equivalent properties
  of materials commonly used in proton dosimetry.},\ }\href@noop {} {\bibfield
  {journal} {\bibinfo  {journal} {Appl. Radiat. Isotopes}\ }\textbf {\bibinfo
  {volume} {83}},\ \bibinfo {pages} {pp. 122} (\bibinfo {year}
  {2013})}\BibitemShut {NoStop}%
\bibitem [{\citenamefont {Söderman}\ \emph {et~al.}( Mar)\citenamefont
  {Söderman}, \citenamefont {Asplund}, \citenamefont {Åse
  Allansdotter~Johnsson}, \citenamefont {Vikgren}, \citenamefont {Norrlund},
  \citenamefont {Molnar}, \citenamefont {Svalkvist}, \citenamefont {Månsson},\
  and\ \citenamefont {Båth}}]{christina2015}%
  \BibitemOpen
  \bibfield  {author} {\bibinfo {author} {\bibfnamefont {C.}~\bibnamefont
  {Söderman}}, \bibinfo {author} {\bibfnamefont {S.}~\bibnamefont {Asplund}},
  \bibinfo {author} {\bibnamefont {Åse Allansdotter~Johnsson}}, \bibinfo
  {author} {\bibfnamefont {J.}~\bibnamefont {Vikgren}}, \bibinfo {author}
  {\bibfnamefont {R.~R.}\ \bibnamefont {Norrlund}}, \bibinfo {author}
  {\bibfnamefont {D.}~\bibnamefont {Molnar}}, \bibinfo {author} {\bibfnamefont
  {A.}~\bibnamefont {Svalkvist}}, \bibinfo {author} {\bibfnamefont {L.~G.}\
  \bibnamefont {Månsson}},\ and\ \bibinfo {author} {\bibfnamefont
  {M.}~\bibnamefont {Båth}},\ }\bibfield  {title} {\bibinfo {title} {Image
  quality dependency on system configuration and tube voltage in chest
  {tomosynthesis—A} visual grading study using an anthropomorphic chest
  phantom},\ }\href@noop {} {\bibfield  {journal} {\bibinfo  {journal} {Med.
  Phys.}\ }\textbf {\bibinfo {volume} {42(3)}},\ \bibinfo {pages} {1200}
  (\bibinfo {year} {2015 Mar})}\BibitemShut {NoStop}%
\bibitem [{\citenamefont {Battaglia}\ \emph {et~al.}(2016)\citenamefont
  {Battaglia}, \citenamefont {Schardt}, \citenamefont {Espino}, \citenamefont
  {Gallardo}, \citenamefont {Cortés-Giraldo}, \citenamefont {Quesada},
  \citenamefont {Lallena}, \citenamefont {Miras},\ and\ \citenamefont
  {Guirado}}]{battaglia2016}%
  \BibitemOpen
  \bibfield  {author} {\bibinfo {author} {\bibfnamefont {M.~C.}\ \bibnamefont
  {Battaglia}}, \bibinfo {author} {\bibfnamefont {D.}~\bibnamefont {Schardt}},
  \bibinfo {author} {\bibfnamefont {J.}~\bibnamefont {Espino}}, \bibinfo
  {author} {\bibfnamefont {M.}~\bibnamefont {Gallardo}}, \bibinfo {author}
  {\bibfnamefont {M.~A.}\ \bibnamefont {Cortés-Giraldo}}, \bibinfo {author}
  {\bibfnamefont {J.~M.}\ \bibnamefont {Quesada}}, \bibinfo {author}
  {\bibfnamefont {A.~M.}\ \bibnamefont {Lallena}}, \bibinfo {author}
  {\bibfnamefont {H.}~\bibnamefont {Miras}},\ and\ \bibinfo {author}
  {\bibfnamefont {D.}~\bibnamefont {Guirado}},\ }\bibfield  {title} {\bibinfo
  {title} {{Dosimetry for low energy protons with ionization chambers and EBT3
  films in the Bragg peak region}},\ }\href@noop {} {\bibfield  {journal}
  {\bibinfo  {journal} {Physica Medica}\ }\textbf {\bibinfo {volume} {32}},\
  \bibinfo {pages} {204} (\bibinfo {year} {2016})}\BibitemShut {NoStop}%
\bibitem [{\citenamefont {Casolaro}\ \emph {et~al.}(ar 9)\citenamefont
  {Casolaro}, \citenamefont {Campajola},\ and\ \citenamefont {{Breglio \emph{et
  al}.}}}]{casolaro2019}%
  \BibitemOpen
  \bibfield  {author} {\bibinfo {author} {\bibfnamefont {P.}~\bibnamefont
  {Casolaro}}, \bibinfo {author} {\bibfnamefont {L.}~\bibnamefont
  {Campajola}},\ and\ \bibinfo {author} {\bibfnamefont {G.}~\bibnamefont
  {{Breglio \emph{et al}.}}},\ }\bibfield  {title} {\bibinfo {title} {Real-time
  dosimetry with radiochromic films},\ }\href@noop {} {\bibfield  {journal}
  {\bibinfo  {journal} {Sci Rep}\ }\textbf {\bibinfo {volume} {9(1)}},\
  \bibinfo {pages} {5307} (\bibinfo {year} {2019 Mar 9})}\BibitemShut {NoStop}%
\bibitem [{\citenamefont {Rezaeian}\ \emph {et~al.}(ar 2)\citenamefont
  {Rezaeian}, \citenamefont {Kashian},\ and\ \citenamefont
  {Mehrara}}]{rezaeian2023}%
  \BibitemOpen
  \bibfield  {author} {\bibinfo {author} {\bibfnamefont {P.}~\bibnamefont
  {Rezaeian}}, \bibinfo {author} {\bibfnamefont {S.}~\bibnamefont {Kashian}},\
  and\ \bibinfo {author} {\bibfnamefont {R.}~\bibnamefont {Mehrara}},\
  }\bibfield  {title} {\bibinfo {title} {Investigation of the effective atomic
  number dependency on kinetic energy using collision stopping powers for
  electrons, protons, alpha, and carbon particles},\ }\href@noop {} {\bibfield
  {journal} {\bibinfo  {journal} {Sci Rep}\ }\textbf {\bibinfo {volume}
  {13(1)}},\ \bibinfo {pages} {3573} (\bibinfo {year} {2023 Mar
  2})}\BibitemShut {NoStop}%
\bibitem [{\citenamefont {Gu}\ \emph {et~al.}(l 21)\citenamefont {Gu},
  \citenamefont {Cunningham}, \citenamefont {Santiburcio}, \citenamefont
  {Pieve}, \citenamefont {Artacho},\ and\ \citenamefont {Kohanoff}}]{gu2020}%
  \BibitemOpen
  \bibfield  {author} {\bibinfo {author} {\bibfnamefont {B.}~\bibnamefont
  {Gu}}, \bibinfo {author} {\bibfnamefont {B.}~\bibnamefont {Cunningham}},
  \bibinfo {author} {\bibfnamefont {D.~M.}\ \bibnamefont {Santiburcio}},
  \bibinfo {author} {\bibfnamefont {F.~D.}\ \bibnamefont {Pieve}}, \bibinfo
  {author} {\bibfnamefont {E.}~\bibnamefont {Artacho}},\ and\ \bibinfo {author}
  {\bibfnamefont {J.}~\bibnamefont {Kohanoff}},\ }\bibfield  {title} {\bibinfo
  {title} {Efficient ab initio calculation of electronic stopping in disordered
  systems via geometry pre-sampling: Application to liquid water},\ }\href@noop
  {} {\bibfield  {journal} {\bibinfo  {journal} {J. Chem. Phys.}\ }\textbf
  {\bibinfo {volume} {153(3)}},\ \bibinfo {pages} {034113} (\bibinfo {year}
  {2020 Jul 21})}\BibitemShut {NoStop}%
\bibitem [{\citenamefont {Kononov}\ \emph {et~al.}(2023)\citenamefont
  {Kononov}, \citenamefont {Hentschel},\ and\ \citenamefont {{Hansen \emph{et
  al}.}}}]{kononov2023}%
  \BibitemOpen
  \bibfield  {author} {\bibinfo {author} {\bibfnamefont {A.}~\bibnamefont
  {Kononov}}, \bibinfo {author} {\bibfnamefont {T.~W.}\ \bibnamefont
  {Hentschel}},\ and\ \bibinfo {author} {\bibfnamefont {S.~B.}\ \bibnamefont
  {{Hansen \emph{et al}.}}},\ }\bibfield  {title} {\bibinfo {title} {Trajectory
  sampling and finite-size effects in first-principles stopping power
  calculations},\ }\href@noop {} {\bibfield  {journal} {\bibinfo  {journal}
  {npj Computational Materials}\ }\textbf {\bibinfo {volume} {9}} (\bibinfo
  {year} {2023})}\BibitemShut {NoStop}%
\bibitem [{\citenamefont {Koval}\ \emph {et~al.}(2012)\citenamefont {Koval},
  \citenamefont {Sánchez-Portal}, \citenamefont {Borisov},\ and\ \citenamefont
  {Muiño}}]{koval12}%
  \BibitemOpen
  \bibfield  {author} {\bibinfo {author} {\bibfnamefont {N.~E.}\ \bibnamefont
  {Koval}}, \bibinfo {author} {\bibfnamefont {D.}~\bibnamefont
  {Sánchez-Portal}}, \bibinfo {author} {\bibfnamefont {A.~G.}\ \bibnamefont
  {Borisov}},\ and\ \bibinfo {author} {\bibfnamefont {R.~D.}\ \bibnamefont
  {Muiño}},\ }\bibfield  {title} {\bibinfo {title} {Dynamic screening of a
  localized hole during photoemission from a metal cluster},\ }\href@noop {}
  {\bibfield  {journal} {\bibinfo  {journal} {Nanoscale Res. Lett.}\ }\textbf
  {\bibinfo {volume} {7}},\ \bibinfo {pages} {447} (\bibinfo {year}
  {2012})}\BibitemShut {NoStop}%
\bibitem [{\citenamefont {Borisov}\ \emph {et~al.}(2004)\citenamefont
  {Borisov}, \citenamefont {Sánchez-Portal}, \citenamefont {Muiño},\ and\
  \citenamefont {Echenique}}]{borisov2004}%
  \BibitemOpen
  \bibfield  {author} {\bibinfo {author} {\bibfnamefont {A.~G.}\ \bibnamefont
  {Borisov}}, \bibinfo {author} {\bibfnamefont {D.}~\bibnamefont
  {Sánchez-Portal}}, \bibinfo {author} {\bibfnamefont {R.~D.}\ \bibnamefont
  {Muiño}},\ and\ \bibinfo {author} {\bibfnamefont {P.~M.}\ \bibnamefont
  {Echenique}},\ }\bibfield  {title} {\bibinfo {title} {Building up the
  screening below the femtosecond scale},\ }\href@noop {} {\bibfield  {journal}
  {\bibinfo  {journal} {Chem. Phys. Lett.}\ }\textbf {\bibinfo {volume}
  {387}},\ \bibinfo {pages} {95} (\bibinfo {year} {2004})}\BibitemShut
  {NoStop}%
\bibitem [{\citenamefont {Quijada}\ \emph {et~al.}(2010)\citenamefont
  {Quijada}, \citenamefont {Muiño}, \citenamefont {Borisov}, \citenamefont
  {Alonso},\ and\ \citenamefont {Echenique}}]{quijada10}%
  \BibitemOpen
  \bibfield  {author} {\bibinfo {author} {\bibfnamefont {M.}~\bibnamefont
  {Quijada}}, \bibinfo {author} {\bibfnamefont {R.~D.}\ \bibnamefont {Muiño}},
  \bibinfo {author} {\bibfnamefont {A.~G.}\ \bibnamefont {Borisov}}, \bibinfo
  {author} {\bibfnamefont {J.~A.}\ \bibnamefont {Alonso}},\ and\ \bibinfo
  {author} {\bibfnamefont {P.~M.}\ \bibnamefont {Echenique}},\ }\bibfield
  {title} {\bibinfo {title} {Lifetime of electronic excitations in metal
  nanoparticles},\ }\href@noop {} {\bibfield  {journal} {\bibinfo  {journal}
  {New J. Phys.}\ }\textbf {\bibinfo {volume} {12}},\ \bibinfo {pages} {053023}
  (\bibinfo {year} {2010})}\BibitemShut {NoStop}%
\bibitem [{\citenamefont {S.~Tanuma}\ and\ \citenamefont
  {Penn}(1994)}]{tanuma1994a}%
  \BibitemOpen
  \bibfield  {author} {\bibinfo {author} {\bibfnamefont {C.~J.~P.}\
  \bibnamefont {S.~Tanuma}}\ and\ \bibinfo {author} {\bibfnamefont {D.~R.}\
  \bibnamefont {Penn}},\ }\bibfield  {title} {\bibinfo {title} {{Calculations
  of Electron Inelastic Mean Free Paths}},\ }\href@noop {} {\bibfield
  {journal} {\bibinfo  {journal} {Surf. Interface Anal.}\ }\textbf {\bibinfo
  {volume} {21}},\ \bibinfo {pages} {165} (\bibinfo {year} {1994})}\BibitemShut
  {NoStop}%
\bibitem [{\citenamefont {Inagaki}\ \emph {et~al.}(1977)\citenamefont
  {Inagaki}, \citenamefont {Arakawa}, \citenamefont {Hamm},\ and\ \citenamefont
  {Williams}}]{inagaki1977}%
  \BibitemOpen
  \bibfield  {author} {\bibinfo {author} {\bibfnamefont {T.}~\bibnamefont
  {Inagaki}}, \bibinfo {author} {\bibfnamefont {E.~T.}\ \bibnamefont
  {Arakawa}}, \bibinfo {author} {\bibfnamefont {R.~N.}\ \bibnamefont {Hamm}},\
  and\ \bibinfo {author} {\bibfnamefont {M.~W.}\ \bibnamefont {Williams}},\
  }\bibfield  {title} {\bibinfo {title} {Optical properties of polystyrene from
  the near-infrared to the x-ray region and convergence of optical sum rules},\
  }\href@noop {} {\bibfield  {journal} {\bibinfo  {journal} {Phys. Rev. B}\
  }\textbf {\bibinfo {volume} {15}},\ \bibinfo {pages} {3243} (\bibinfo {year}
  {1977})}\BibitemShut {NoStop}%
\bibitem [{\citenamefont {de~Vera}\ \emph {et~al.}(2011)\citenamefont
  {de~Vera}, \citenamefont {Abril},\ and\ \citenamefont
  {Garcia-Molina}}]{deVera2011}%
  \BibitemOpen
  \bibfield  {author} {\bibinfo {author} {\bibfnamefont {P.}~\bibnamefont
  {de~Vera}}, \bibinfo {author} {\bibfnamefont {I.}~\bibnamefont {Abril}},\
  and\ \bibinfo {author} {\bibfnamefont {R.}~\bibnamefont {Garcia-Molina}},\
  }\bibfield  {title} {\bibinfo {title} {Inelastic scattering of electron and
  light ion beams in organic polymers},\ }\href@noop {} {\bibfield  {journal}
  {\bibinfo  {journal} {J. Appl. Phys.}\ }\textbf {\bibinfo {volume} {109}},\
  \bibinfo {pages} {094901} (\bibinfo {year} {2011})}\BibitemShut {NoStop}%
\bibitem [{\citenamefont {ICRU49}(1993)}]{icru49:1993}%
  \BibitemOpen
  \bibfield  {author} {\bibinfo {author} {\bibnamefont {ICRU49}},\ }\bibfield
  {title} {\bibinfo {title} {{ICRU Report 49}: {Stopping Powers and Ranges for
  Protons and Alpha Particles}},\ }\href@noop {} {\bibfield  {journal}
  {\bibinfo  {journal} {Journal of the ICRU}\ }\textbf {\bibinfo {volume}
  {os-25}} (\bibinfo {year} {1993})}\BibitemShut {NoStop}%
\bibitem [{\citenamefont {ICRU37}(1984)}]{icru37:1984}%
  \BibitemOpen
  \bibfield  {author} {\bibinfo {author} {\bibnamefont {ICRU37}},\ }\bibfield
  {title} {\bibinfo {title} {{ICRU Report 37: Stopping Powers for Electroncs
  and Positrons}},\ }\href@noop {} {\bibfield  {journal} {\bibinfo  {journal}
  {Journal of the ICRU}\ }\textbf {\bibinfo {volume} {os-19}} (\bibinfo {year}
  {1984})}\BibitemShut {NoStop}%
\bibitem [{\citenamefont {Ziegler}\ \emph {et~al.}(2010)\citenamefont
  {Ziegler}, \citenamefont {Ziegler},\ and\ \citenamefont
  {Biersack}}]{srim2010}%
  \BibitemOpen
  \bibfield  {author} {\bibinfo {author} {\bibfnamefont {J.~F.}\ \bibnamefont
  {Ziegler}}, \bibinfo {author} {\bibfnamefont {M.}~\bibnamefont {Ziegler}},\
  and\ \bibinfo {author} {\bibfnamefont {J.}~\bibnamefont {Biersack}},\
  }\bibfield  {title} {\bibinfo {title} {Srim – the stopping and range of
  ions in matter (2010)},\ }\href@noop {} {\bibfield  {journal} {\bibinfo
  {journal} {Nuclear Instruments and Methods in Physics Research Section B:
  Beam Interactions with Materials and Atoms}\ }\textbf {\bibinfo {volume}
  {268}},\ \bibinfo {pages} {1818} (\bibinfo {year} {2010})}\BibitemShut
  {NoStop}%
\bibitem [{\citenamefont {Paul}(2016)}]{iaea}%
  \BibitemOpen
  \bibfield  {author} {\bibinfo {author} {\bibfnamefont {H.}~\bibnamefont
  {Paul}},\ }\href@noop {} {\bibinfo {title} {Electronic stopping power of
  matter for ions}},\ \bibinfo {howpublished} {\emph{available from}
  \url{https://www-nds.iaea.org/stopping/}} (\bibinfo {year}
  {2016})\BibitemShut {NoStop}%
\bibitem [{\citenamefont {Garcia-Molina}\ \emph {et~al.}(2014)\citenamefont
  {Garcia-Molina}, \citenamefont {Abril}, \citenamefont {de~Vera},
  \citenamefont {Kyriakou},\ and\ \citenamefont {Emfietzoglou}}]{molina2013}%
  \BibitemOpen
  \bibfield  {author} {\bibinfo {author} {\bibfnamefont {R.}~\bibnamefont
  {Garcia-Molina}}, \bibinfo {author} {\bibfnamefont {I.}~\bibnamefont
  {Abril}}, \bibinfo {author} {\bibfnamefont {P.}~\bibnamefont {de~Vera}},
  \bibinfo {author} {\bibfnamefont {I.}~\bibnamefont {Kyriakou}},\ and\
  \bibinfo {author} {\bibfnamefont {D.}~\bibnamefont {Emfietzoglou}},\
  }\bibfield  {title} {\bibinfo {title} {A study of the energy deposition
  proﬁle of proton beams in materials of hadron therapeutic interest},\
  }\href@noop {} {\bibfield  {journal} {\bibinfo  {journal} {Appl. Radiat.
  Isotopes}\ }\textbf {\bibinfo {volume} {83}},\ \bibinfo {pages} {pp. 109}
  (\bibinfo {year} {2014})}\BibitemShut {NoStop}%
\bibitem [{\citenamefont {Dalton}\ and\ \citenamefont {Turner}(
  Sep)}]{dalton1968}%
  \BibitemOpen
  \bibfield  {author} {\bibinfo {author} {\bibfnamefont {P.}~\bibnamefont
  {Dalton}}\ and\ \bibinfo {author} {\bibfnamefont {J.~E.}\ \bibnamefont
  {Turner}},\ }\bibfield  {title} {\bibinfo {title} {New evaluation of mean
  excitation energies for use in radiation dosimetry},\ }\href@noop {}
  {\bibfield  {journal} {\bibinfo  {journal} {Health Phys.}\ }\textbf {\bibinfo
  {volume} {15(3)}},\ \bibinfo {pages} {257} (\bibinfo {year} {1968
  Sep})}\BibitemShut {NoStop}%
\end{thebibliography}%

\end{document}